\documentclass[
nofootinbib,
amsmath,amssymb,
aps, physrev,
superscriptaddress
]{revtex4-2}

\usepackage{array}
\usepackage{graphicx}
\usepackage{dcolumn}
\usepackage{bm}
\usepackage{mathtools}

\usepackage[caption=false]{subfig}
\usepackage{multirow}
\usepackage{xcolor}

\usepackage{hyperref}

\newcommand\norm[1]{\left\lVert#1\right\rVert}
\newcommand{\pC}[2]{\parbox[c]{#1}{\centering #2}}
\newcommand{\pL}[2]{\parbox[c]{#1}{\raggedright #2}}
\newcounter{tab_row}
\newcommand*\thetabrow{C\arabic{tab_row}}

\newcommand{\subfigref}[2]{%
  \hyperref[#1]{\ref*{#1}#2}%
}

\begin{document}

\preprint{APS/123-QED}

\title{\textbf{
Optimizing Superconducting Microwave Cavities for Gravitational Wave Sensing
}}%

\author{Lars Fischer}
\altaffiliation[Present address: ]{Institute for Quantum Electronics \& Quantum Center, ETH Zurich, 8093 Zurich, Switzerland}
\affiliation{Universität Hamburg,
Luruper Chaussee 149, 22761, Hamburg, Germany}

\author{Wolfgang Hillert}%
\affiliation{Universität Hamburg,
Luruper Chaussee 149, 22761, Hamburg, Germany}

\author{Tom Krokotsch}%
\email{Contact author: tom.krokotsch@desy.de}
\affiliation{Universität Hamburg,
Luruper Chaussee 149, 22761, Hamburg, Germany}

\author{Gudrid Moortgat-Pick}
\affiliation{Universität Hamburg,
Luruper Chaussee 149, 22761, Hamburg, Germany}
\affiliation{Deutsches Elektronen-Synchrotron DESY,
 Notkestraße 85, 22607, Hamburg, Germany}

\begin{abstract}
Superconducting microwave cavities loaded with radio frequency fields are a powerful tool to search for weak forces or electromagnetic perturbations due to new physics. One source of such signals can be high-frequency gravitational waves emitted from cosmological or unknown astrophysical events. However, large sensitivity improvements are still necessary to reach the parameter space motivated by theoretical models. In this work, we present a formalism to guide the design of such detectors across a broad range of frequencies and signal forms. By incorporating the interdependencies of all relevant parameters and the back-action of the electromagnetic fields on the cavity structure, we describe how figures of merit can be derived for a broad class of relevant experimental setups. Using a two-dimensional model, we demonstrate this formalism and present examples for optimized cavity geometries. We find that the choice of geometry alone can increase the signal-to-noise power ratio by an order of magnitude compared to an existing prototype of the same size. We also show that different physics goals lead to different optimal cavities, thus proving the need to consider such figures of merit at an early stage when designing a new detector.
\end{abstract}

\maketitle
\tableofcontents


%
\section{Introduction}
The current state-of-the-art technology for gravitational wave (GW) detectors are optical cavities, resonantly excited with high-power lasers \cite{First_GW_Observation}. However, superconducting radio-frequency (SRF) cavities with a driven eigenmode are emerging as a promising alternative for sensing weak electromagnetic (EM) perturbations. While SRF cavity detectors are expected to suffer from larger noise than optical cavities for gravitational wave frequencies $f_g\lesssim 10\,\text{kHz}$, they are among the most competitive technologies to search for \emph{high-frequency gravitational waves} (HFGWs) at GW frequencies $f_g\gtrsim 10\,\text{kHz}$ \cite{berlin_mago20_2023}. HFGWs are a largely unexplored probe of new physics and have been receiving increasing attention in recent years \cite{aggarwal2025challengesopportunitiesgravitationalwave}. Combined experimental and theoretical efforts show a promising avenue for tests of a first SRF cavity prototype for GW detection in the near future \cite{berlin_mago20_2023, fischer2024characterisationmagocavitysuperconducting, dokuyucu2026cryogenicrfcharacterizationmago}.
At the same time, similar SRF cavity-based detectors are being developed to search for axion dark matter \cite{giaccone_design_2022, Slac_heterodyne, SHANHE, Navarro-Madrid:2026fql} and generally have simultaneous, but limited, sensitivity to HFGWs. Unfortunately, this first generation of SRF cavity detectors will only be able to search for HFGW sources located close to Earth \cite{Berlin_narrowing_down, aggarwal2025challengesopportunitiesgravitationalwave}, and major improvements in sensitivity are needed to probe HFGWs of cosmological origin \cite{dagnolo2025classicalandquantumheuristics, aggarwal2025challengesopportunitiesgravitationalwave, Gaudio:2026bhe}. Furthermore, many orders of magnitude in frequency must be covered, which requires several new detectors.

However, so far, there has been no guidance on how to design SRF cavities \emph{optimized} for HFGW searches. Although the sensitivity can always be improved by increasing the size of any cavity, a significant sensitivity enhancement is also possible by choosing an optimal cavity geometry for a fixed size.

In this work, we develop and apply a formalism for a systematic search of optimal microwave cavity geometries in different GW frequency ranges. 
Most existing SRF cavities have been developed for the well-known application of accelerating charged particles in particle accelerators. For example, `TESLA'-shaped accelerator cavities were designed by choosing a cylinder with rounded edges as a `baseline' design and optimizing the curvature profile to maximize the accelerating field and quality factor while minimizing sensitivity to instabilities and imperfections \cite{TESLA_cavities, TESLA_design}. 

The conceptual challenge for SRF cavities as GW detectors is that finding an optimal geometry requires considering a much larger class of shapes, as any change to the cavity walls affects the EM properties of the cavity and the GW coupling at the same time. 
We propose beginning the development process by narrowing down potential geometries that feature the best possible theoretical coupling to identify a `baseline' design. Then, starting from this baseline design, more technical design requirements can be investigated which are relevant for an experimental operation. The goal of this work is to find such a smaller class of baseline geometries. Therefore, we present here the first step in an optimization chain that requires additional optimization steps after a suitable class of shapes has been found. Due to the unique symmetries and distinct interactions of gravitational waves with EM cavities, optimal geometries are expected to differ notably from cavities optimized for EM signals from dark matter \cite{ADMX, HAYSTAC, Slac_heterodyne, giaccone_design_2022} and from accelerator cavities. 

\begin{figure}[t!]
    \centering
    \includegraphics[width=\linewidth]{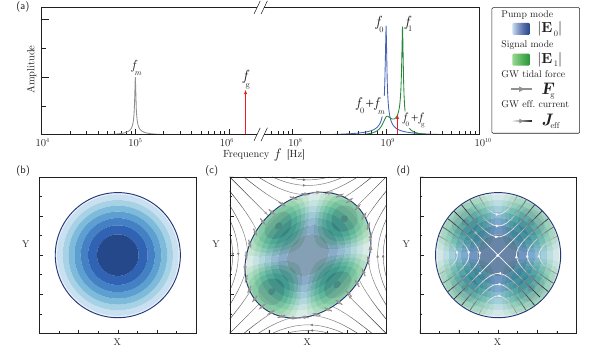}
    \caption{\textbf{Schematic of GWs coupling to driven microwave cavities.} (a) Shows the different frequencies at play while a GW-induced EM signal is generated in a typical SRF cavity. The cavity is driven at frequency $f_0$ and a signal sideband due to a GW with frequency $f_g$ appears at $f = f_0 + f_\text{g}$ within the bandwidth of the signal mode $f_1$. An excitation at a mechanical resonance frequency $f_m=f_g$ would also be resonantly enhanced. In (b) a cross-section of a cylindrical cavity driven in the TM$_{010}$ mode is shown, with the electric field strength in blue. (c) illustrates the mechanical wall deformation caused by the GW tidal force $\bm{F}_\text{g}$, modifying the boundary conditions and coupling energy into the signal mode, with its electric field shown in green. In (d) the electromagnetic coupling is illustrated, where the GW induces an effective current $\bm{J}_\text{eff}$ from the pump field, which excites the signal mode, shown in green. Stronger current strengths are indicated by a darker gray shading of the lines.}
    \label{fig:SRF_GW_Illustration}
\end{figure}

The general principle of GW detection with driven microwave cavities is illustrated in Figure \ref{fig:SRF_GW_Illustration}. A resonant \emph{pump} mode of the cavity is excited with oscillating EM fields at a  frequency $f_0$. A GW parametrically perturbs the cavity and converts a fraction of the stored energy to the sideband frequencies $f_0\pm f_\text{g}$. This frequency conversion can follow from a mechanical cavity deformation or a direct perturbation of the electromagnetic field, sometimes referred to as `(inverse) Gertsenshtein effect' \cite{gertsensthein}. If the cavity contains another \emph{signal} mode with resonance frequency at $f_1=f_0+ f_\text{g}$, it amplifies the up-converted signal, proportional to a quality factor $Q_1$, where values $Q_1\gtrsim10^{10}$ are achieved routinely in modern SRF cavities \cite{padamsee}. The frequency spacing of the resonant modes in SRF cavities can be adjusted to resonantly enhance GW signals across a broad range from kHz to $\sim\text{GHz}$ frequencies. Key advantages of SRF cavities include their compact design, decades of development in particle accelerator technology, and access to existing infrastructure at accelerator facilities worldwide. Additionally, SRF cavities can store significantly more electromagnetic energy densities than optical cavities and can operate with near-quantum-limited noise performance to sense weak electromagnetic signals \cite{fischer2024characterisationmagocavitysuperconducting, giaccone_design_2022, SHANHE, Slac_heterodyne, DarkSRF}.

We begin by introducing a formalism to analyze the effect of a cavity's shape on the GW-induced signal in chapter \ref{sec:mech_and_EM_interactions}. By redefining overlap coefficients from the literature, we can reveal qualitative geometrical features that are relevant for sensitivity improvements. Afterward, we express the signal and noise powers in terms of our new coupling coefficients. Chapter \ref{sec:optimizing_the_geometry} describes the construction of the figures of merit for a numerical optimization for different gravitational wave frequency ranges and signal analysis methods from the formalism introduced previously. Referring to the figures of merit, we draw general lessons on the role different cavity parameters play in the optimization of the geometry. The analysis is backed up by appendices \ref{sec:mech_res} and \ref{sec:FOMs}, which provide expressions from which figures of merit can be constructed for all relevant frequency regimes and modes of operation, taking back-action of the electromagnetic fields on the cavity walls fully into account.

As a demonstration, we implement figures of merit in two dimensions to numerically perform the optimization process for two examples in chapter \ref{sec:2D_optimization}. We compare the resulting sensitivity with the farthest developed prototype for an SRF-GW detector to date, the `MAGO' cavity \cite{ballantini_microwave_2005,fischer2024characterisationmagocavitysuperconducting}, and show that the ratio of signal and noise power can increase by more than an order of magnitude for a detector of the same size. 

\section{Gravitational wave signals in loaded cavities}\label{sec:mech_and_EM_interactions}
Before we can define the figures of merit, we first establish a formalism for analyzing the signal and noise powers in terms of dimensionless and normalized coupling coefficients. 
In this section we therefore revise the equations needed to describe the mechanical and electromagnetic interactions of a GW with microwave cavities. We carefully separate the efficiency of the GW energy transfer in the form of coupling coefficients from other experimental parameters and the dimensions of the detector. This allows us to identify the key aspects by which a detector can be optimized.
For details on the different derivations, we refer to \cite{Gue:2026kga}.

\subsection{Signal power}
First, we will express the GW-induced signal power in terms of power spectral densities (PSDs) $S_f(\omega)$ defined through $\langle f(\omega)f(\omega')\rangle_\text{ens}=\delta(\omega-\omega')S_f(\omega)$, where $f(t)$ is a (stochastic) function and $\langle\cdot\rangle_\text{ens}$ is an ensemble average so that the time-averaged squared amplitude becomes $\langle f(t)^2\rangle=(2\pi)^{-2}\int_{-\infty}^\infty d\omega\,S_f(\omega)$. If $f(t)$ is chosen so that $\langle f(t)^2\rangle$ is the signal or noise power received by the readout system, we refer to the PSD as $S_\text{sig}(\omega)$ or $S_\text{noise}(\omega)$ respectively. However, it will often be more instructive to express noise in terms of the false GW strain one would infer from measuring it. 

For this purpose, we write the spatial components of a monochromatic GW metric perturbation in transverse-traceless (TT) gauge  as
\begin{equation}\label{eq:mochoromatic_GW_TT_definition}
    h^\text{TT}_{ij}=h(t)\,e^{-i\bm{k}_g\cdot\bm{x}}\hat{h}_{ij}= h(t)e^{-i\bm{k}_g\cdot\bm{x}}\left[\cos{2\psi}\,(t_i^1t_j^1-t_i^2t_j^2)+\sin{2\psi}\,(t_i^1t_j^2+t_i^2t_j^1)\right]\,,
\end{equation}
where $h(t)=h_0 e^{i\omega_g t}$ denotes the amplitude, $\psi$ the polarization angle of the wave, $\hat{\bm{k}}_g=\bm{k}_g\,c/\omega_g=(\sin{\theta}\cos{\varphi},\sin{\theta}\sin{\varphi}, \cos{\theta})$ the direction of the GW, and two transverse vectors, which can be chosen as $\bm{t}^1=(-\sin{\varphi}, \cos{\varphi},0)$ and $\bm{t}^2=(\cos{\theta}\cos{\varphi},\cos{\theta}\sin{\varphi}, -\sin{\theta})$. Furthermore, we will also use the expression $\hat{\bm{h}}=(\hat{h}_{ij})_{i,j\in\{1,2,3\}}$ to denote a matrix containing the spatial components of the normalized strain metric.

This allows us to define the  \emph{signal transfer function} as the ratio between the signal PSD and the PSD of the GW strain amplitude 
\begin{equation}
    T_\text{sig}(\omega)=\frac{S_\text{sig}(\omega)}{S_h(\omega-\omega_0)}\,.
\end{equation}
The noise PSD divided by the transfer function can then be used to define the \emph{strain-equivalent noise amplitude spectral density}
\begin{equation}\label{eq:strain_equivalent_noise_PSD}
    \sqrt{S_h^\text{noise}(\omega)}\coloneqq\sqrt{\frac{S_\text{noise}(\omega)}{T_\text{sig}(\omega)}}\,,
\end{equation}
which is a measure of the minimal detectable GW strain \cite{maggiore_gravitational_2008}. 

Depending on the GW frequency range, we will use different coordinate systems (or `gauges') to express the GW perturbation, as it can simplify the resulting PSDs in different cases. The first gauge are \emph{proper detector} (PD) coordinates, which can be understood as measuring distances using rigid rulers, which are not themselves perturbed by the GW. At \emph{long-wavelengths} $\omega_g L_\text{cavity}\ll c$ with $L_\text{cavity}^2=\max_{\bm{x}\in V}{\bm{x}^2}$, the GW acts like a force density in PD coordinates and the signal in PD coordinates arises predominantly due to the mechanical deformation of the cavity walls.  At higher frequencies, the signal contribution includes mechanical \emph{and} EM contributions and becomes harder to compute \cite{Gue:2026kga}, especially for non-monochromatic GWs \cite{Fischer:2025mpz}. 

However, if the cavity walls are approximately free falling, i.e. elastic forces in the cavity walls are negligible compared to the PD GW force, the signal in \emph{transverse-traceless} (TT) coordinates arises predominantly due to the EM perturbation of the pump mode \cite{Gue:2026kga}. This happens, since TT coordinates measure distances using rulers that are freely falling themselves, and deform along with the cavity. Therefore, the displacement of the cavity walls in TT gauge is negligible in free fall $|\delta\bm{x}_\text{wall}^\text{TT}|\ll h_0L_\text{cavity}$. 

The free falling approximation is expected to be valid at frequencies above all dominant mechanical resonances, which can still occur within the long-wavelength approximation. This allows for a frequency window in which the EM signal in TT gauge is equal to the mechanical signal in PD gauge. 

The signal transfer function using the PD gauge on the signal mode resonance is then given by \cite{Gue:2026kga}
\begin{equation}\label{eq:mechanical_signal_PSD}
    T_\text{sig}^\text{mech.}(\omega_1)=\frac{Q_1^2}{32\mu_0}\frac{\omega_1}{Q_\text{cpl}}\frac{B_0^2}{V}\left|\sum_m\frac{\omega_g^2\int_{V_\text{w}}dV\frac{\rho}{M}\bm{\xi}_m\cdot\hat{\bm{h}}\cdot\bm{x}}{\omega_m^2-\omega_g^2+i\frac{\omega_g\omega_m}{Q_m}}\int d\bm{A}\cdot\bm{\xi}_m\left(\bm{B}_1\cdot\bm{B}_0^*-\bm{E}_1\cdot\bm{E}_0^*\right)\right|^2\,,
\end{equation}
and the EM signal transfer function using TT gauge on resonance by\footnote{The volume integral appearing here can be derived from $(j_n^\text{bulk})^\text{TT}$ in reference \cite{Gue:2026kga} by using Maxwell's equations, integrating by parts, and using the boundary condition of $\bm{E}_1$.}
\begin{align}\label{eq:em_sig_psd}
    T_\text{sig}^\text{EM}(\omega_1)=&\frac{Q_1^2}{8\mu_0}\frac{\omega_1}{Q_\text{cpl}}\frac{B_0^2}{V}\Bigg|\hat{h}_{ij}\int dVe^{-i\bm{k}_g\cdot\bm{x}}\Big((B_1^*)^iB_0^j+(E_1^*)^iE_0^j+\epsilon_{ikl}\frac{k_g^l}{\omega_1}\left((E_1^*)^kB_0^j-(E_1^*)^jB_0^k\right)\Big)\Bigg|^2\,.
\end{align}
$\bm{E}_n(\bm{x})$ and $\bm{B}_n(\bm{x})$ are the electric and magnetic field distributions of the EM eigenmodes of the cavity with perfectly conducting walls and are orthonormal $(\bm{E}_n, \bm{E}_m)_V=\delta_{nm}$ under the inner product $ (\bm{a},\bm{b})_V=\frac{1}{V}\int_{V}dV\bm{a}(\bm{x})^\ast\cdot\bm{b}(\bm{x})$.  $\bm{\xi}_m(\bm{x})$ are the displacement fields of the mechanical eigenmodes of the cavity walls with mass density $\rho(\bm{x})$ and total mass $M=\int_{V_\text{w}} dV\,\rho$, and are also orthonormal $(\bm{\xi}_m,\bm{\xi}_l)_\text{M}=\delta_{ml}$ under the inner product $(\bm{a},\bm{b})_\text{M}=\frac{1}{M}\int_{V_\text{w}}dV\rho(\bm{x})\,\bm{a}(\bm{x})\cdot\bm{b}(\bm{x})$. Therefore, in our notation, the eigenmode distributions $\bm{E}_n,\bm{B}_n,\bm{\xi}_m$ are \emph{dimensionless}. The pump and signal modes are denoted by the indices $n=0$ and $n=1$ respectively. $\omega_1=2\pi f_1$ and $\omega_m=2\pi f_m$ are the angular signal and mechanical resonance frequencies. $B_0$ is the (dimensionful) amplitude with which the pump mode is excited so that the pump EM fields in the cavity are given by
\begin{subequations}
    \begin{align}
        \bm{B}_\text{pump}(\bm{x},t)&=B_0\text{Re}\left[e^{i\omega_0t}\bm{B}_0(\bm{x})\right]\,,\\
        \bm{E}_\text{pump}(\bm{x},t)&=c\,B_0\text{Re}\left[e^{i\omega_0t}\bm{E}_0(\bm{x})\right]\,.
    \end{align}
\end{subequations}
$Q_m$ is the quality factor of the mechanical mode and $Q_1=(1/Q_\text{int}+1/Q_\text{cpl})^{-1}$ is the \emph{loaded} quality factor of the signal mode, meaning it includes both internal losses through the cavity walls $\propto 1/Q_\text{int}$ as well as losses into the signal readout $\propto 1/Q_\text{cpl}$.\footnote{If the pump mode is not driven with the same antenna used for the readout, the signal mode can experience further losses. However, we will assume that the pump antenna only has negligible coupling to the readout, which is possible even when the pump and signal mode geometries are identical \cite{bernard_rf_2000}.} In the following, we will often use the coupling parameter $\beta\coloneqq Q_\text{int}/Q_\text{cpl}$ to characterize the readout. The cases for $\beta$ that we will consider are \emph{critical coupling} $\beta\simeq1$ and \emph{overcoupling} $\beta>1$.

Importantly, the result above is only valid when the EM forces on the cavity walls generated by the pump and signal fields are negligible. Such back-action effects often need to be accounted for in the vicinity of mechanical resonances and change the resulting signal power. A detailed description of the effect is given in appendix \ref{sec:mech_res} and its consequences on the optimization are explained in appendix \ref{sec:FOMs}.

\subsection{Coupling coefficients}\label{sec:coupling_coefficients}
To identify the combinations of parameters that lead to the best GW sensitivity, it is useful to separate the pump field magnitude and cavity size from the dimensionless efficiency with which GW energy is being transferred to the signal readout. To this end, we define coupling coefficients, which only depend on the geometry of the cavity walls, as well as the electromagnetic and mechanical eigenmode distributions. For the coefficients, we choose different normalizations from previous works, which help to reveal geometric features of the cavity and eigenmodes that improve the GW sensitivity.

The efficiency by which the GW mechanically generates an electromagnetic signal in the cavity is characterized by two different coupling coefficients. First, the set of $\Gamma_m^{ij}$ which measure the overlap of the GW tidal force in the direction $j$ and the displacement field of the $m$th mechanical mode in the direction $i$. Second, the set of $C_{n_0n_1}^m$ which measure the coupling strength between the EM modes $n_0$ and $n_1$ induced by the vibrational mode $m$. The GW-mechanical coupling efficiency can be read off equation \eqref{eq:mechanical_signal_PSD} and is given by
\begin{equation}\label{eq:GW_mech_coupling_def}
    \Gamma^{ij}_m=\frac{(\xi_m^i,x^j)_\text{M}}{R_\text{eff}}\,,
\end{equation}
with the \emph{effective radius} $R_\text{eff}=\norm{\bm{x}}_\text{M}\eqqcolon \sqrt{(\bm{x},\bm{x})_\text{M}}$. The combination of coefficients entering the signal power for a given GW can then be obtained from
\begin{equation}
    \Gamma_m\coloneqq \hat{h}_{ij}\Gamma^{ij}_m\,.
\end{equation}
Dividing by the effective radius in equation \eqref{eq:GW_mech_coupling_def} differs from the typical factor $V^{1/3}$ in the previous literature \cite{berlin_mago20_2023, lowenberg_lorentz_2023} and has the benefit of enforcing a strict normalization.
This can be seen using the Cauchy-Schwarz inequality $0\leq|\Gamma_m|\leq \lVert\hat{\bm{h}}\cdot\bm{x}\rVert_\text{M}/\lVert\bm{x}\rVert_\text{M}\leq1$ where the largest coupling is achieved in the case $\bm{\xi}_m\propto\hat{\bm{h}}\cdot\bm{x}$.

In order to define the mechanical-EM coupling, we introduce another inner product 
\begin{equation}
    (\bm{a},\bm{b})_S\coloneqq \frac{1}{A_S}\int_{\partial V}dS\,\bm{a}(\bm{x})^\ast\cdot\bm{b}(\bm{x})\,,
\end{equation}
where $\partial V$ is the inner boundary of the cavity walls with surface area $A_S$ so that we can write 
\begin{equation}\label{eq:C01_Definition}
    C_{01}^m\coloneqq\frac12\frac{\big(\xi_m^n\bm{B}_0, \bm{B}_1\big)_S-
    (\xi_m^n\bm{E}_0, \bm{E}_1)_S}{\norm{\bm{B}_0}_S\norm{\bm{B}_1}_S}\,,
\end{equation}
where $\xi_m^n=\bm{\xi}_m\cdot\hat{\bm{n}}_S$ is the displacement field of a mechanical mode projected onto the cavity surface normal $\hat{\bm{n}}_S$ and $\norm{\cdot}_S=\sqrt{(\cdot,\cdot)_S}\,$. Since the overlap integrals in the numerator only depend on the surface fields, we chose to normalize the coupling by the \emph{surface} field averages, which differs from previous treatments \cite{ballantini_microwave_2005, berlin_mago20_2023, lowenberg_lorentz_2023}.
This coefficient has no strict upper bound but in typical cases is $C_{01}^m\lesssim1$. 

In the free falling regime, the GW force dominates over all elastic forces, and the dependence on the mechanical resonance frequencies in equation \eqref{eq:mechanical_signal_PSD} disappears $\omega_g^2/(\omega_m^2-\omega_g^2+i\omega_g\omega_m/Q_m)\to-1$ at GW frequencies above all strongly coupled mechanical modes. Since the signal no longer depends on the parameters $\omega_m$ and $Q_m$, the total GW-EM coupling is simply determined by the sum $\sum_m\Gamma_mC_{01}^m$. We can evaluate the sum by using the completeness of the mechanical eigenmode basis
\begin{equation}
\sum_m\bm{\xi}^m\Gamma_m=\sum_m\bm{\xi}^m(\bm{\xi}_m\bm{,\hat{h}}\cdot\bm{x})_\text{M}/\norm{\bm{x}}_\text{M}=\bm{\hat{h}}\cdot\bm{x}/\norm{\bm{x}}_\text{M}\,,
\end{equation} 
and therefore simplify the sum 
\begin{equation}\label{eq:C01_full_mech_sum}
    C_{01}^g\coloneqq \sum_mC_{01}^m\Gamma_m=\frac12\frac{\big((\hat{\bm{n}}_S\cdot\hat{\bm{h}}\cdot\bm{x})\bm{B}_0, \bm{B}_1\big)_S-((\hat{\bm{n}}_S\cdot\hat{\bm{h}}\cdot\bm{x})\bm{E}_0, \bm{E}_1)_S}{\norm{\bm{B}_0}_S\norm{\bm{B}_1}_S\norm{\bm{x}}_\text{M}}\,,
\end{equation}
which yields a single coupling coefficient for the mechanical coupling of the GW to the signal mode.  

Last, we introduce the direct coupling between a gravitational wave and electromagnetic fields $\kappa_{01}$ from equation \eqref{eq:em_sig_psd} with $\omega_g=\omega_1-\omega_0$
\begin{equation} \label{eq:directCouplingCoefficient}
    \kappa_{01}\coloneqq \hat{h}_{ij}\kappa_{01}^{ij}\,,
\end{equation} 
with the response matrix
\begin{equation}
    \kappa_{01}^{ij}=\frac{1}{V}\int_VdV\,e^{-i\bm{k}_g\cdot\bm{x}}\left[(B_1^*)^iB_0^j+(E_1^*)^iE_0^j+\frac{\omega_1-\omega_0}{\omega_1}\epsilon_{ikl}\hat{k}_g^l\left((E_1^*)^kB_0^j-(E_1^*)^jB_0^k\right)\right]\,.
\end{equation}

Using these coefficients, we can finally rewrite the signal transfer functions on resonance $\omega_g=\omega_1-\omega_0$ in equations \eqref{eq:mechanical_signal_PSD} and \eqref{eq:em_sig_psd}
\begin{align}\label{eq:mechanical_signal_PSD_with_couplings}
T_\text{sig}^\text{mech.}(\omega_1)&=\frac{\omega_1}{8\mu_0}\frac{Q_\text{int}\beta}{(1+\beta)^2}\frac{(R_\text{eff}A_S)^2}{V}B_c^2\frac{\norm{\bm{B}_0}_S^2\norm{\bm{B}_1}_S^2}{\underset{\bm{x}\in\partial V}{\max}|\bm{B}_0(\bm{x})|^2}\left|\sum_m\frac{\omega_g^2\,\Gamma_mC_{01}^m}{\omega_m^2-\omega_g^2+i\frac{\omega_g\omega_m}{Q_m}}\right|^2\,, \\
   T_\text{sig}^\text{EM}(\omega_1) &= \frac{\omega_1}{8\mu_0}\frac{Q_\text{int}\beta}{(1+\beta)^2} V\frac{B_c^2}{\underset{\bm{x}\in\partial V}{\max}|\bm{B}_0(\bm{x})|^2} | \kappa_{01}|^2\,,     \label{eq:EM_signal_PSD_with_couplings}
\end{align}
where we have chosen the pump amplitude as
\begin{equation}
    B_0=\frac{B_c}{\underset{\bm{x}\in\partial V}{\max}|\bm{B}_0(\bm{x})|}\,,
\end{equation}
which is the largest possible value for $B_0$ until the maximal surface field reaches the critical field $B_c$, above which the cavity loses its superconducting properties.
The improved normalizations of the coefficients reveals important subtleties in the dependencies on various cavity parameters and will be analyzed in the following chapter \ref{sec:optimizing_the_geometry}.

The case of unloaded cavities with a strong static background magnetic field $\bm{B}_0^\text{static}=(\bm{B}_0^\text{static})^*$ can be described with the same equations above by setting $\omega_0=0$ and replacing $\bm{B}_0\to2\bm{B}_0^\text{static}$ to account for down-converted frequency components of the pump field, which were neglected for an oscillating background.

\subsection{Noise power}\label{sec:signal_and_noise_PSDs}
In order to discuss optimal cavity geometries, it is not sufficient to consider only the signal response in equations \eqref{eq:EM_signal_PSD_with_couplings}, but we need to consider signal-to-noise ratios. One irreducible noise source is given by thermal excitations of the cavity fields with the PSD
\begin{equation}\label{eq:thermal_noise_PSD}
    S_\text{th. em.}(\omega)=4\pi k_BT_c\begin{cases}\frac{\beta}{(1+\beta)^2},\quad&\omega\simeq\omega_1-\omega_0\\\frac{1}{Q_\text{int}Q_\text{cpl}}\left(\frac{\omega_1}{\omega}\right)^2,\quad&\omega_0+\omega\gg\omega_1+\frac{\omega_1}{Q_1}\end{cases}\,,
\end{equation}
where $T_c$ is the temperature of the cavity walls. The readout system at a possibly different temperature $T_r$ sources broadband thermal noise measured in reflection 
\begin{equation}\label{eq:readout_noise_PSD}
    S_r(\omega)=\pi\hbar\omega\,n_T^r\,,
\end{equation} 
where $n_T^r=\left(e^{\hbar\omega/k_BT_r}-1\right)^{-1}+1$ is the number of photons per unit bandwidth in the readout system. The first term describes thermally created photons, which can be reduced to zero at temperatures $k_BT_r\ll\hbar\omega$. The second term accounts for half a photon per unit bandwidth from zero point fluctuations and another half of photon's energy from a quantum-limited phase-insensitive amplifier \cite{caves_quantum_1982, Clerk_Introduction}.

The last noise sources we consider are thermal excitations of a mechanical mode due to the effective external force PSD \cite{maggiore_gravitational_2008}
\begin{equation}\label{eq:thermal_mechanical_noise_PSD}
S_{f_m^\text{th}}(\omega)=4\pi k_BT_cM\frac{\omega_m}{Q_m}\,. 
\end{equation}

Although this list does not encompass all potential noise mechanisms, the included sources represent the primary limitations within the relevant parameter regime. Effects like RF power leaking from the driven pump mode into the signal readout or other excitations of the cavity vibrations are expected to be reducible below the thermal and quantum noise floors. This can be achieved with a suitable RF system and mechanical isolation of the cavity as described in references \cite{pegoraro_operation_1978, ballantini_microwave_2005}. In particular, for frequencies $f_1-f_0\simeq f_\text{g}\gg\text{kHz}$ where the signal and pump modes are well separated and less external noise forces are expected, the technical complexity is reduced.

More details on the resulting noise spectrum can be found in appendix \ref{sec:mech_res}, but can be summarized as follows. Noise from thermal vibrations dominates near mechanical resonances when $\omega_g\simeq\omega_m=\omega_1-\omega_0$. Away from any resonance, readout noise power dominates and stays constant while the signal power falls off $\propto\omega^{-4}$. On an electromagnetic resonance $\omega_g\simeq\omega_1-\omega_0$, thermal emissions can also dominate. However, thermal emissions can always be tuned to be subdominant to readout noise by increasing $\beta$, since $S_\text{th. em}/S_\text{r}\propto\frac{\beta}{(1+\beta)^2}$ but $T_\text{sig}/S_\text{th. em}$ does not depend on $\beta$.

\section{Geometrical figures of merit}\label{sec:optimizing_the_geometry}
In the previous chapter, we derived the most important properties of an ideal SRF cavity for GW detection. The resulting transfer functions in equations \eqref{eq:mechanical_signal_PSD_with_couplings} and \eqref{eq:EM_signal_PSD_with_couplings} reveal several important features that were not obvious in previous analyses \cite{ballantini_microwave_2005, berlin_mago20_2023, lowenberg_lorentz_2023}. For example, the mechanical signal does not scale with the total energy stored in the pump mode and can even be lowered by increasing the cavity volume. 

Here we describe how those results can be used to identify geometrical features of SRF cavities, which can improve the GW sensitivity. The combination of cavity parameters that are maximized in a cavity optimization depends on the targeted GW sources and frequency range, the signal analysis procedure, and the dominant noise in the detector. Instead of listing all possible combinations, we will analyse optimal cavity features only conceptually in this chapter and describe how they can be turned into figures of merit (FOMs) for optimization in practice. A full list of quantities that could enter possible FOMs in different experimental setups is given in appendix \ref{sec:FOMs}. Often, this requires choosing an optimal value for the antenna coupling $Q_\text{cpl}$ to the signal mode, in particular when back-action of the EM fields on the cavity walls is significant, which is discussed in detail in appendix \ref{sec:mech_res}.

The FOMs also need to take inter-dependencies of quantities in our PSDs into account which we have not yet made explicit. The quality factor for internal losses $Q_\text{int}$ is determined by the signal mode distribution \cite{padamsee} 
\begin{equation}\label{eq:Qint_definition}
 Q_\text{int}=\mu_0\frac{\omega_1}{R_S(\omega_1)}\frac{V}{A_S}\frac{1}{\norm{\bm{B}_1}_S^2} \,,  
\end{equation} 
where we assumed a constant RF surface resistance $R_S$ across the walls. In general $R_S$ is frequency dependent and is usually decomposed as $R_S = R_\text{BCS}+ R_\text{res}$ with a constant, $R_\text{res} \gtrsim \mathcal{O}(\text{n}\Omega)$ residual resistance\footnote{Residual resistances of $3-5\,\text{n}\Omega$ were achieved with cavities produced in larger scales for accelerators \cite{PhysRevAccelBeams.20.042004, gonnella:srf2019-frcaa3}, while individual cavities with special treatments have shown values as low as $1-2\,\text{n}\Omega$ \cite{padamsee}.}, and a frequency-dependent BCS resistance, 
\begin{equation}\label{eq:BCS_Resistance}
    R_\text{BCS}(\omega_1) \simeq 2.7\,\text{n}\Omega \,\left(\frac{1.8\,\text{K}}{T}\right)\left(\frac{f_1}{\text{GHz}}\right)^2e^{-\left(\frac{\Delta(T)}{k_BT}-\frac{1.76\,T_c}{1.8\,\text{K}}\right)}\eqqcolon \rho_\text{BCS}^\text{GHz}\,\left(\frac{f_1}{\text{GHz}}\right)^2\,,
\end{equation} for state-of-the-art niobium cavities \cite{LILJE2004213}, where $\Delta(T)\approx1.76\,k_BT_c$ is the energy gap of the superconductor with critical temperature $T_c$.

We can construct our FOMs as a quantity proportional to the signal-to-noise ratio (SNR) for the targeted GW source. The key ingredients are usually the dominant noise strain on resonance $S_h^\text{noise}(\omega_1)$ and the bandwidth $\text{BW}$ for which $S_h^\text{noise}(\omega)\approx\text{const.}\,$.\footnote{In most cases $\text{BW}\sim\omega_1/Q_\text{cpl}$, however when resonant noise sources dominate outside the cavity bandwidth or mechanical resonances enter, BW can be unrelated to $Q_\text{cpl}$ as further discussed in appendix \ref{sec:FOMs}.}  For example, the optimal SNR for a coherent GW, whose waveform is known theoretically, can be obtained by using \emph{matched filtering} \cite{maggiore_gravitational_2008}
\begin{equation}\label{eq:SNR_matched_filter}
    \text{SNR}^\text{linear}=\sqrt{2\pi\int_{-\infty}^\infty df\frac{|h(f)|^2}{S_{h}^\text{noise}(f)}}\propto\sqrt{\text{BW}/S_{h}^\text{noise}(f_1)}\,,
\end{equation}
which is only possible to compute for a \emph{linear readout} when the signal power is analyzed along with all phase information of the sideband oscillation. When only time-averaged power spectra have been recorded, the SNR for a matched filter becomes \cite{berlin_heterodyne_2021, aggarwal2025challengesopportunitiesgravitationalwave}
\begin{equation}\label{eq:PSD_SNR}
    \text{SNR}^\text{quadratic}=\sqrt{t_\text{int}\int df\frac{S_\text{sig}(f)^2}{S_\text{noise}(f)^2}}\propto\sqrt{\text{BW}}/S_{h}^\text{noise}(f_1)\,.
\end{equation}
Clearly, the bandwidth and resonant strain sensitivity enter the SNR in different combinations, depending on the readout strategy. Even more combinations are possible for different GW types, analysis strategies, and if e.g. multiple detectors are cross-correlated  \cite{maggiore_gravitational_2008, aggarwal2025challengesopportunitiesgravitationalwave}. Furthermore, the FOM can either be chosen as $\text{FOM}\sim\text{SNR}$, or as a quantity derived from the SNR, such as the scanning rate $R_\text{scan}\propto(\text{SNR}^\text{quadratic})^2$ \cite{malnou_squeezed_2019} in the search for a narrowband signal by tuning the signal resonance. We will discuss examples for both cases in section \ref{sec:2D_optimization}. In any case, the resulting  FOM can typically be approximated as a product of sensitivity bandwidth and resonant strain sensitivity raised to different powers. Therefore, to keep our analysis as general as possible, we only provide the strain sensitivity and bandwidth in appendix \ref{sec:FOMs} for all relevant parameter regions, which can be combined into FOMs as necessary.

Even though different FOMs can be considered for a cavity optimization, we can nevertheless recognize common features which improve SNRs like the ones in equations \eqref{eq:SNR_matched_filter} and \eqref{eq:PSD_SNR}. However, we need to distinguish between different GW frequency regimes.

\subsection{Long-wavelength regime}
In the long-wavelength limit  $L_\text{cavity}f_g\ll c$, the dominant signal can be described in PD coordinates as emerging due to the GW interaction with the cavity walls and the EM surface fields. This may also include frequencies where the cavity walls are freely falling, and we can express the signal transfer function using either the volume integral in equation \eqref{eq:em_sig_psd} or the surface integral in equation \eqref{eq:mechanical_signal_PSD}.\footnote{$S_\text{sig}^\text{mech.}\approx S_\text{sig}^\text{EM}$ can be shown explicitly in free fall by using Maxwell's equations and the EM boundary conditions in the limit $\omega_g\ll\omega_0$ and $\omega_1-\omega_0\ll\omega_0$.}  However, we will use the surface integral form and thus equation \eqref{eq:mechanical_signal_PSD_with_couplings}, as it provides a better qualitative understanding of the optimal cavity wall and EM mode geometries. Since the sensitivity is always determined from the signal relative to noise, we need to distinguish the results depending on which noise source dominates. To back up claims on the sensitivity scaling with different parameters, we will also give references to results from appendix \ref{sec:FOMs} in tables \ref{tab:FOM_Collection} and \ref{tab:FOM_scaling}, which can be ignored without affecting the qualitative arguments of this section.

\paragraph{GW tidal force $\bm{F}_\text{g}$:}
The tidal force from a GW grows with the effective radius $R_\text{eff}$ of the cavity. Since \emph{no} noise source increases with $R_\text{eff}$, it is guaranteed to improve the sensitivity. Importantly, $R_\text{eff}$ describes the average spatial extent of the cavity walls, and \emph{not} the volume. For example, the tidal force of a GW on an ellipsoid is larger than that of a sphere with the same volume. 

While the GW force also increases with the mass $M$ of the cavity walls, it does not affect the signal power. However, since the force of thermal vibrations in equation \eqref{eq:thermal_mechanical_noise_PSD} (the dominant noise on mechanical resonance) scales with $\sqrt{M}$, a large cavity mass can increase the SNR. Since a large mass also suppresses vibrations induced from external forces, it is in principle \emph{always} beneficial to increase the cavity mass as much as possible by e.g. building thick walls. 
\paragraph{Cavity volume:}
The mechanical transfer function in equation \eqref{eq:mechanical_signal_PSD_with_couplings} increases with the effective radius, due to the tidal force and with the cavity surface area, along which the wall movement couples to the EM fields. However, notably, the cavity volume \emph{decreases} the transfer function. This arises since a larger cavity means the GW signal also needs to excite a larger volume to reach the same signal power. Even though for typical cavities the parameters are linked $V\sim R_\text{eff}A$, a cavity with maximized signal transfer function is expected to have reduced volume while still maintaining a large effective radius and coupling area.
\paragraph{Mode geometry:}\label{sec:optimizing_geometry_field_flatness}
The next aspect to consider is the geometry of the electromagnetic modes. Because the GW couples to the signal mode through the walls, \emph{only} the fields of the EM modes at the surface affect the sensitivity, as equation \eqref{eq:mechanical_signal_PSD} shows. The maximal magnetic field on the surface cannot exceed the critical field $B_c$ of the superconducting material. Therefore, a large ratio $\norm{\bm{B}_0}_S/\underset{\bm{x}\in \partial V}{\max}|\bm{B}_0(\bm{x})|\leq1$ is optimal. Constant magnetic fields on the surface would be an ideal pump mode and are, for instance, nearly achieved by the TM$_{010}$ mode in cylinders. 

For the signal mode, it can be beneficial to store most of its energy near the surface walls $\norm{\bm{B}_1}_S\gg1$ and avoid large fields within the bulk. While this increases the coupling to the pump mode via wall vibrations, it also leads to higher surface losses and thus a lower $Q_\text{int}$. Therefore, when $Q_\text{int}$ enters the SNR, the coefficient $\norm{\bm{B}_1}_S$ can cancel. This happens when thermal emissions are the dominant noise source (see rows \ref{R3} and \ref{R5} in table \ref{tab:FOM_Collection}). 

\paragraph{The signal frequency $\omega_1$:}
The role of the signal frequency in the sensitivity depends strongly on the specific FOM which is demonstrated in tables \ref{tab:FOM_Collection} and \ref{tab:FOM_scaling} in the appendix. Whenever the surface resistance $R_S(\omega)$ enters the SNR, it converges for large frequencies $\sqrt{\text{SNR}}\sim\omega_1/\sqrt{R_S(\omega_1)}\rightarrow 2\pi\,\text{GHz}/\sqrt{\rho_\text{BCS}^\text{GHz}}$ when thermal emissions dominate on resonance (row \ref{R3} in table \ref{tab:FOM_Collection}). When readout noise dominates on resonance and the readout is critically coupled, the SNR is maximized for a unique frequency $\omega_1=\sqrt{R_\text{res}/ \rho_\text{BCS}^\text{GHz}}\,\text{GHz}$ (see row \ref{R5} in table \ref{tab:FOM_Collection}). When the resonant bandwidth $\omega_1/Q_1$ is part of the optimization, it adds further benefits for increasing $\omega_1$.

\paragraph{Coupling area $A_S$ and volume efficiency:}
The last aspect to consider is the efficiency with which a cavity design uses available space in a cryostat and how much cavity area can be used to couple to a GW. As a general rule, the cavity should have a large area where the GW force is large, i.e.\ far away from the cavity center of mass, and also support its strongest EM fields there. This is often not the case for cavities like spheres or cylinders which efficiently fill out the volume of standard cryostats. Therefore, a compromise between efficient use of cryostat space and maximal GW coupling is usually optimal.

Overall, we expect that using more sophisticated cavity geometries can combine the benefits of the different design aspects. Depending on the relevant FOM, this includes EM field flatness on the surface and a wide extent of the cavity with a large area where the GW can couple to the EM fields. Furthermore, the mode geometries must match the GW symmetries to have an efficient coupling according to the $C_{01}^m$ coefficient. 

\subsection{Wavelengths comparable to detector size}
At frequencies above the long-wavelength limit, the surface integral in equation \eqref{eq:mechanical_signal_PSD} no longer describes the dominant signal contribution and the direct GW-EM coupling needs to be taken into account as well. As the cavity walls are expected to be freely falling at these frequencies, the easiest way is to use the EM current in TT coordinates as in equation \eqref{eq:em_sig_psd}. Since it cannot be cast into a surface integral at $\omega_gL_\text{cavity}\gtrsim c$ anymore, the scaling with cavity parameters also changes.
From the signal transfer function in equation \eqref{eq:EM_signal_PSD_with_couplings} we can determine the scaling of the direct coupling to the signal field.

\paragraph{Cavity surface and volume:} The signal transfer function in equation \eqref{eq:EM_signal_PSD_with_couplings} increases linearly with the cavity volume, while the noise sources are volume independent. Consequently, a larger volume always improves the sensitivity, as opposed to the case at long wavelengths. For a critically coupled detector readout (row \ref{R7}, table \ref{tab:FOM_Collection}), the SNR scales with $V/\sqrt{A_S}$ and the surface is ideally small to maximize the internal quality factor. 

\paragraph{Mode geometry:}
Like for long wavelengths, the sensitivity can be increased by choosing a pump mode with low $\underset{\bm{x}\in \partial V}{\max}|\bm{B}_0(\bm{x})|$. However, contrary to long wavelengths, there is never a benefit in choosing pump or signal modes with most of their fields near the surface. When the SNR depends on the internal quality factor, it can even be beneficial to have a signal mode with $\norm{\bm{B}_1}_S\ll1$, to reduce thermal dissipation in the walls.

\paragraph{Signal mode frequencies:} 
As for the mechanical interaction, the thermal noise-dominated SNR (row \ref{R7}, table \ref{tab:FOM_Collection}) becomes frequency independent as soon as the BCS resistance dominates. Since the mode frequencies are parametrically connected to the cavity volume via $\omega_{1}\sim V_\text{cav}^{1/3}$, the volume usually cannot be increased simultaneously with $\omega_1$ in general, however the factor $V\omega_1$ still grows. This tradeoff can be circumvented by e.g. using multiple smaller, coupled resonator structures as described in \cite{withers_beehive_2024}.

\section{Cavity optimization in two dimensions}\label{sec:2D_optimization}

\begin{figure}[hb]
    \centering
    \includegraphics[width=\textwidth]{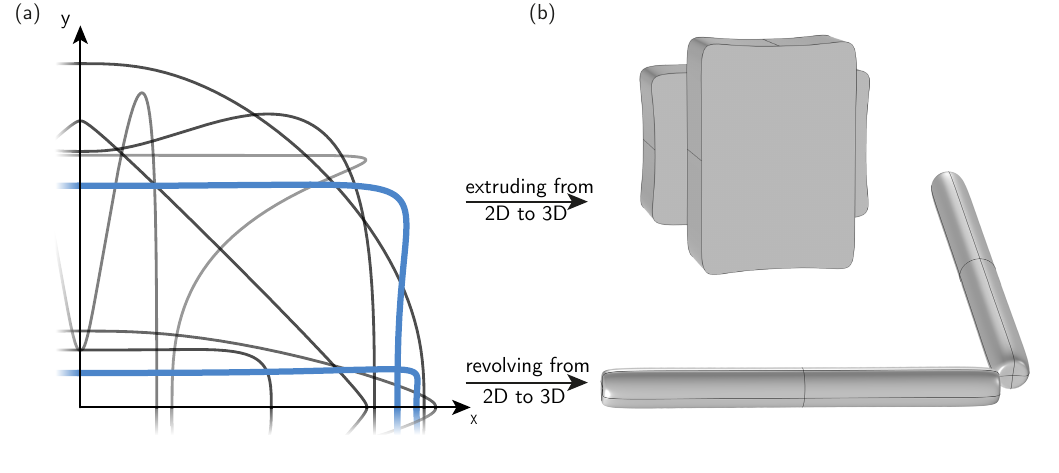}
    \caption{\textbf{Cavity boundary parametrization and extrapolation to 3D} (a) Examples for the different curves $\bm{C}(s)$ for cavity boundaries considered in the optimization. The curves are mirrored into the remaining coordinate quadrants to form a closed boundary, indicated by the extensions beyond the $x-$ and $y-$axis. (b) The two curves highlighted in blue in (a) are extrapolated to a 3D geometry of a two-cell cavity. In the lower cavity, the two copies of the 2D surface have been revolved around their long symmetry axis and placed at a right angle to each other. In the example above, two 2D surfaces have been extruded along their surface normal and placed at a right angle behind each other.}
    \label{fig:two_D_to_three_D}
\end{figure}

To demonstrate the use of FOMs and showcase advantageous geometrical features, we will discuss an optimization for a cavity in two dimensions (2D). Starting a cavity optimization in 2D has the advantage of significantly reduced computational cost, which allows considering larger classes of topologies than in three dimensions. Furthermore, a 2D design can easily be extended into three dimensions by extruding or revolving the 2D design as illustrated in figure \ref{fig:two_D_to_three_D}. 

For our example, we will optimize a cavity for GW frequencies in the long-wavelength approximation with approximately freely falling walls. For typical parameters, this corresponds to the range $100\,\text{kHz}<f_g\ll 1\,\text{GHz}$ \cite{Gue:2026kga}. We will imagine a cavity design where two identical coupled cavity cells are oriented at a right angle and pump and signal mode are chosen to be symmetric and antisymmetric oscillations of the same mode distribution. Such a design provides two nearly degenerate EM modes for the pump and signal fields and enables rejecting the pump in the readout by subtracting the signal from both cells \cite{ballantini_microwave_2005, berlin_mago20_2023}. Furthermore, we will distinguish between the case where the readout system is overcoupled $Q_\text{cpl}\ll Q_\text{int}$ so that $S_r(\omega)\gg S_\text{th. em.}(\omega)$ to achieve a broad resonance (case \eqref{R6} in table \ref{tab:FOM_Collection}) and the case with maximal resonant sensitivity, where $Q_1^\text{cpl}\approx Q_1^\text{int}$, so that $S_\text{th. em.}(\omega)$ is the limiting noise source on resonance (case \eqref{R3} in table \ref{tab:FOM_Collection}). 

For the signal analysis procedure in the broadband case, we will assume a linear readout with matched filtering, which leads us to define the figure of merit $\text{FOM}_\text{broad.}\sim\text{SNR}^\text{linear}$ from equation \eqref{eq:SNR_matched_filter} and using table \ref{tab:FOM_Collection}
\begin{equation}\label{eq:broad_FOM}
    \text{FOM}_\text{broad.}=\frac{\omega_1V_\text{bdy}^{1/3}}{c}\frac{\left|\int d\bm{A}\cdot\hat{\bm{h}}\cdot\bm{x}\left(\bm{B}_0^2-\bm{E}_0^2\right)\right|^2}{V_\text{bdy}V\underset{\bm{x}\in\partial V}{\max}{|\bm{B}_0|^2}}\,.
\end{equation}
This is the combination of all geometry dependent parameters that must be maximized. It has been normalized by the volume available to the detector $V_\text{bdy}$ (e.g., the volume of the helium vessel in a cryostat) to make the FOM dimensionless and independent of the cavity dimensions.

In the case of a resonant scheme, we are assuming a setup where the resonant frequency is changed to scan a larger frequency range. Therefore, we will choose our dimensionless FOM proportional to the scan rate as defined in chapter \ref{sec:optimizing_the_geometry} such that $\text{FOM}_\text{res. scan.}\sim \sqrt{R_\text{scan}}$
\begin{equation}\label{eq:res_FOM}
    \text{FOM}_\text{res. scan.}=\left(\frac{\omega_1V_\text{bdy}^{1/3}}{c}\right)^2\sqrt{\frac{V}{V^{1/3}_\text{bdy}\int dA\,\bm{B}_0^2}}\frac{\left|\int d\bm{A}\cdot\hat{\bm{h}}\cdot\bm{x}\left(\bm{B}_0^2-\bm{E}_0^2\right)\right|^2}{V_\text{bdy}V\underset{\bm{x}\in\partial V}{\max}{|\bm{B}_0|^2}}\,.
\end{equation}
For the maximization of the FOMs, an iteration over as many different types of cavity geometries as possible is necessary. A common approach is to start with a baseline design such as a cylinder or cube and apply different small shape deformations. However, since we are interested in finding such a baseline design in the first place, we instead fully parametrize the cavity boundary as a curve of polynomial functions. We choose to only parametrize one quarter of the cavity as $\bm{C}:[0,1]\to\mathbb{R}^+\times\mathbb{R}^+$, $\bm{C}(s)=(C_x(s),C_y(s))$, and rotate it into the three remaining quadrants of $\mathbb{R}^2$ so that the cavity is mirror symmetric along both coordinate axes. Due to the quadrupole symmetry of the GW, we do not expect asymmetric cavities to have advantages. We will choose the parametrization
\begin{equation}\label{eq:wall_parametrization}
    \bm{C}(s)=\begin{pmatrix}C_x(1)(10s^3 - 15s^4 + 6s^5)+C_x'(0)\left(s - 6s^3 + 8s^4 - 3s^5\right) \\ C_y(0)(1 - 10s^3 + 15s^4 - 6s^5)+C_y'(1)(-4s^3 + 7s^4 - 3s^5)\end{pmatrix}\,,
\end{equation}
which is the unique quintic polynomial expansion fulfilling the smoothness constraints $C_x(0)=C_y(1)=C'_y(0)=C'_x(1)=0$, as well as $\bm{C}''(0)=\bm{C}''(1)=0$ for a smooth curvature. Examples of cavity geometries this parametrization includes are shown in figure \subfigref{fig:two_D_to_three_D}{a}. Taking $C_x(1)$ as the length scale of the cavity, this leaves three dimensionless parameters $C_y(0)/C_x(1)$, $C_x'(0)/C_x(1)$, $C_y'(1)/C_x(1)$ to vary freely. For the boundary constraint from, e.g. a cryostat, we will consider the two cases of circular and rectangular domains. A circular domain models the space constraints in typical vertical cryostats for cavity testing \cite{PhysRevAccelBeams.20.042004}, where the two orthogonal 2D cavity cells can be extended to 3D by extrusion and then placed behind each other. A rectangular domain models the space constraints in typical horizontal cryomodules used in particle accelerators, where we imagine two perpendicular cryomodules, filled with one cavity cell each. In this case, the 2D geometry is then extended to 3D by a revolution around the horizontal axis. These two designs are illustrated in figure \ref{fig:two_D_to_three_D}.

Our optimization procedure consists of obtaining the EM fields and frequencies of the $10$ eigenmodes with lowest frequency of $\sim 10^3$ different combinations of the three free parameters using COMSOL Multiphysics\textsuperscript{\textregistered} \cite{comsol64}. In post-processing, we increase our dataset by rescaling all resulting quantities with different values of $C_x(1)$ and compute the different resulting FOMs for a plus and cross polarized GW perpendicular to the orthogonal cavity cells. Finally, we select the combination of $C_x(1)$, $C_y(0)$, $C_x'(0)$, $C_y'(1)$ and the eigenmode number $n$ that maximizes the sum of the plus and cross polarized FOMs. Therefore, we only optimize for the polarization average of the optimal GW direction. For a wider field of view, an optimization for a full sky average would also be possible, but this requires specifying the cavity proportions in 3D. However, since cavities built from two orthogonal cells inherently support one optimal GW direction with weak angular dependence for other incidence angles, we opt to maximize only with respect to the optimal GW direction. To support this claim, we prove in appendix \ref{sec:Rectangular_overlaps} that the sky-averaged sensitivity of two orthogonal rectangular 3D cavities is simultaneously maximized when only the coupling to the optimal GW incidence angle is maximized.

\begin{figure}[hb]
    \centering
        \includegraphics[width=\textwidth]{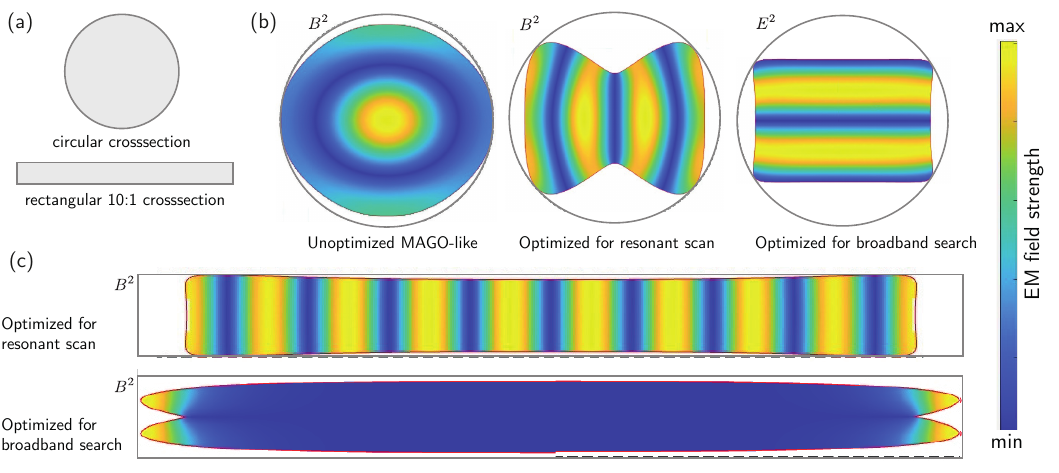}
    \caption{\textbf{Optimization result for a circular and rectangular cross section.} In (a) the cross sections that are used for the optimization results are shown. (b) shows the non-optimal reference geometry of a TE$_{011}$-type mode in a geometry similar to a cell of the  MAGO cavity \cite{ballantini_microwave_2005} and the optimal mode geometries for the case of resonant and broadband scanning of a circular cross section. In (c) the optimal mode geometries for the rectangular cross section are shown. The resonant scanning FOM in equation \eqref{eq:res_FOM} and the broadband FOM in equation \eqref{eq:broad_FOM} were maximized for each cross section. 
    The upper geometry in (c) also corresponds to the optimal geometry for the broadband FOM if the additional constraint $C_y'(1)/C_x(1)<-0.1$ is imposed to avoid the sharp cavity feature in the geometry below. The magnetic or electric field strength of the resonant mode is shown depending on which contributes most to the overlap integral in the FOM.}
    \label{fig:combined_optimization}
\end{figure}

The result of this optimization is shown in figure \ref{fig:combined_optimization} for the circular boundary and a rectangular boundary with an aspect ratio 10, respectively. The aspect ratio is chosen similar to the dimensions of the helium tank in common cryomodules housing two 1.3 GHz TESLA nine-cell cavities \cite{TEICHERT2006239, Stengler:2020vrh, Kaya:2019mrd}. For each geometry, we show the magnitude of the electric or magnetic field, depending on which contributes the most to the overlap integral in equations \eqref{eq:broad_FOM} and \eqref{eq:res_FOM}. As expected, we find that different FOMs lead to different optimal geometries. However, common features are that large surface areas are preferred over large cavity volumes as deduced from our coupling coefficients in chapter \ref{sec:optimizing_the_geometry}. Furthermore, as predicted, modes with little spatial variation (but not necessarily lowest in frequency) are beneficial, and large surface field values where the GW force is strongest are optimal. To quantify the possible improvement in sensitivity due to the choice of cavity geometry, we compare the result with a mode shape similar to the MAGO cavity \cite{ballantini_microwave_2005, fischer2024characterisationmagocavitysuperconducting} as a reference. The corresponding geometry in our 2D parametrization is illustrated in figure \subfigref{fig:combined_optimization}{b} on the left, for which we find  $\text{FOM}^\text{sphere}_\text{res. scan.}=6.0$ and $\text{FOM}^\text{sphere}_\text{broad.}=1.6$.
\begin{table}
    \centering
    \begin{tabular}{|c||c|c|c|}
    \hline
         &Circular Boundary& Rectangular Boundary\\
         \hline\hline
         FOM$^\text{max.}_\text{broad.}/\text{FOM}_\text{broad.}^\text{sphere}$&7.1 &44.9\;(9.8) \\
         \hline
         FOM$^\text{max.}_\text{res. scan.}/\text{FOM}^\text{sphere}_\text{res. scan.}$&7.3 &9.2 \\
         \hline
    \end{tabular}
    \caption{Values for the maximal FOMs corresponding to the geometries in figure \ref{fig:combined_optimization}. The value in brackets corresponds to the optimization with the additional constraint $C_y'(1)/C_x(1)<-0.1$ to avoid a sharp feature. We report the FOMS relative to the values for a MAGO-like geometry, which are $\text{FOM}^\text{sphere}_\text{res. scan.}=6.0$ and $\text{FOM}^\text{sphere}_\text{broad.}=1.6$.}
    \label{tab:optimization_FOM_values}
\end{table}
The FOM values for the different geometries are summarized in table \ref{tab:optimization_FOM_values}. This illustrates that the choice of cavity geometry and eigenmode alone can increase FOMs by factors $\sim 10$, which corresponds to an improvement in GW strain sensitivity by a factor of $\sim 3$ with respect to the MAGO design. Since the FOM values are independent of the cavity length scale, further sensitivity improvements are possible by building cavities that fill a larger volume $V_\text{bdy}$. The broadband FOM for the cavity in figure \subfigref{fig:combined_optimization}{c} at the bottom shows the biggest improvement over MAGO, which is due to the fact that the magnetic fields are concentrated exclusively near the end walls, where the signal emerges. However, the sharp tip promotes field emission of electrons, which likely limits the performance of the cavity\footnote{The boundary conditions on the curvature of the parametrization in equation \eqref{eq:wall_parametrization} discourage sharp features, but curves as in figure \ref{fig:combined_optimization} are still possible and are smooth everywhere.}. 
The more practical design in the upper panel of figure \subfigref{fig:combined_optimization}{c} yields a smaller FOM but is optimal for resonant scanning \emph{and} is also among the best geometries for broadband searches in our optimization.

Furthermore, we see that the maximal FOMs in an elongated rectangular boundary are larger than the optimal FOMs in a circular boundary. This suggests that cavities in two orthogonal horizontal cryomodules offer more GW sensitivity per cryostat volume than two stacked cavity cells in a vertical cryostat. However, this comparison is only based on the GW sensitivity obtained with a given cryostat volume and ignores several practical matters which must be considered in any experimental setup. This includes limitations due to field emissions, multipacting, fabrication constraints, cavity stability, and sensitivity to fabrication imperfections. 

While we have only focused on two specific detector setups in this chapter, several more types of optimization would be possible (see table \ref{tab:FOM_Collection} in appendix \ref{sec:FOMs}) and are expected to yield different cavity designs. For example, when the GW wavelength becomes comparable to the size of the detector, we expect cavities with maximal volume such as cylinders to be optimal, due to the different volume scaling discussed in section \ref{sec:optimizing_the_geometry}. If the detector is designed to be excited near mechanical resonances, the optimization has to take more parameters into account. This includes the material and thickness of the cavity walls and mechanical boundary conditions.
\section{Conclusion and outlook}
In this work, we have described how to find a baseline design for an SRF cavity optimized for high-frequency GW detection. The developed formalism identifies geometric features that can be tuned to enhance the signal transfer function by e.g. minimizing the cavity volume while maximizing the coupling area and effective radius.

Our numerical results show that optimizing the geometry can improve the GW strain sensitivity by more than a factor of three beyond what is achieved by the MAGO cavity as a first prototype. However, future HFGW detectors still require significantly better signal transfer functions to reach the cosmologically relevant parameter space \cite{dagnolo2025classicalandquantumheuristics}. One way to achieve this is by increasing the detector size, which can dramatically improve the strain sensitivity. Furthermore, future breakthroughs in SRF technology could increase the critical magnetic fields and allow larger pump amplitudes. 

Another promising avenue is the operation of the readout system beyond the standard quantum limit. Analogous axion searches have already reached significant milestones by using single photon detectors \cite{axion_spd_detector}, vacuum squeezing \cite{malnou_squeezed_2019}, and quantum back-action-free signal amplification  \cite{CASEFIRE_Experiment_Demonstration}. Employing these techniques in SRF-GW detectors would improve their bandwidths and increase their sensitivity to any kind of broadband signal. 

However, most importantly, a next generation of cavity-based GW detectors must first maximize the capabilities of existing SRF technology through the choice of cavity geometry. This will pave the way to probe uncharted aspects of the universe in the future.

\subsection*{Acknowledgments}
The authors thank Giovanni Marconato, Krisztian Peters, and Marc Wenskat for valuable discussions. WH, GMP, and TK acknowledge support by the Deutsche Forschungsgemeinschaft (DFG, German Research Foundation) under Germany's Excellence Strategy - EXC 2121 `Quantum Universe' - 390833306.

\appendix
\renewcommand\thefigure{A\arabic{figure}} 
\renewcommand\thetable{A\arabic{table}}
\setcounter{figure}{0}
\setcounter{table}{0}
\section{Sensitivity in the presence of mechanical back-action}\label{sec:mech_res}
In this appendix, we discuss the effect that EM back-action has on the cavity walls of resonantly driven SRF cavities. Further, we discuss ways to maximize sensitivity in the presence of back-action by choosing an optimal antenna coupling $Q_\text{cpl}$ to the signal mode. The optimal coupling values are an important input for constructing FOMs that take all parameter dependencies into account.
\subsection{On mechanical resonance}
It seems promising not only to use the electromagnetic resonator to enhance the GW signal but also to resonantly excite a mechanical vibration of the cavity walls. This can be achieved in case the frequency difference of the pump and signal modes have been tuned to match a mechanical resonant mode that can be efficiently excited by a GW $\omega_g=\omega_m=\omega_1-\omega_0$. However, in that case, the forces from the EM fields on the cavity walls cannot be neglected \cite{maggiore_gravitational_2008, lowenberg_lorentz_2023}. The coupled equations of motion for the amplitude of the signal mode $b_1(t)$ and of a mechanical eigenmode $q_m(t)$ are given by \cite{Gue:2026kga}
\begin{align}\label{eq:EOM_b_with_BA_short}
    \ddot{b}_1+\frac{\omega_1}{Q_1}\dot{b}_1+\omega_1^2b_1&=\gamma_1q_me^{i\omega_0t}+\omega_1^2b_\text{in}\,,\\\label{eq:EOM_q_with_BA_short}
    \ddot{q}_m+\frac{\omega_m}{Q_m}\dot{q}_m+\omega_m^2q_m&=\frac{1}{M}f^\text{ext}_m+\gamma_m b_1e^{-i\omega_0t}\,,
\end{align}
where we have introduced a possible input $b_\text{in}$ for noise entering the signal mode directly, we further kept the external force driving the cavity walls general, and defined $\gamma_1\coloneqq \omega_1^2\frac{\norm{\bm{B}_0}_S\norm{\bm{B}_1}_S}{V}A_SC_{01}^mB_0$, $\gamma_m\coloneqq\frac{1}{M}\frac{\norm{\bm{B}_0}_S\norm{\bm{B}_1}_S}{\mu_0}A_S(C_{01}^m)^*B_0$. The solution in frequency space for an up-converted signal is given by
\begin{align}\label{eq:b_1_solution_with_damp}
    b_1(\omega_0+\omega)=\frac{\frac{\gamma_1}{M}f^\text{ext}_m(\omega)+\omega_1^2R_m(\omega)\,b_\text{in}(\omega_0+\omega)}{R_1(\omega_0+\omega)R_m(\omega)-\gamma_1\gamma_m}\,,
\end{align}
with the resonance functions $R_i(\omega)\coloneqq\omega_i^2-\omega^2+i\frac{\omega\omega_i}{Q_i}$. The term $\gamma_1\gamma_m$ in the denominator is a result from the back-action of the fields on the wall and can significantly change the transfer function of the detector. Notably, it also affects fields that enter the EM resonator directly through $b_\text{in}$ such as thermal noise. We can see from equation \eqref{eq:b_1_solution_with_damp} that back-action is negligible if
\begin{equation}\label{eq:back_action_criterion}
    \frac{\gamma_1\gamma_m}{|R_1(\omega_0+\omega)R_m(\omega)|}\ll1\,.
\end{equation}
For large back-action, any mechanical and electromagnetic mode pair can form a coupled system with a resonance for both modes oscillating in phase or with a $\pi$ phase difference. These split resonances can in principle favorably widen the bandwidth of the cavity but are not enhanced by the full $Q_1Q_m$ quality of both resonators. This can be seen in figure \ref{fig:mechanical_resonance_sensitivity} on the top left, where the readout noise strain for $Q_\text{cpl}=10^{10}$ splits into two resonances, which merge into one peak for lower quality factors.  The signal power on resonance $\omega_1-\omega_0=\omega_m$ can be  maximized under the condition
\begin{equation}\label{eq:max_m_res_Q_cpl}
    Q_\text{cpl}^\text{max. m. res.}=\left(\frac{1}{Q_\text{int}}+\frac{Q_m}{Q_\text{BA}}\right)^{-1}\,,
\end{equation}
where we have introduced a back-action damping quality factor 
\begin{equation}\label{eq:Q_back_action}
Q_\text{BA}\coloneqq\frac{\omega_1^2(\omega_1-\omega_0)^2}{\gamma_1\gamma_m}=1.2\cdot10^8\frac{M}{10^3\,\text{kg}}\left(\frac{\omega_1-\omega_0}{10\,\text{kHz}}\right)^2\frac{V/{1\,\text{m}^3}}{(A_S/{1\,\text{m}^2})^2}\left(\frac{0.1\,\text{T}}{B_0}\frac{1}{\norm{\bm{B}_1}_S\norm{\bm{B}_0}_SC_{01}^m}\right)^2\,.  
\end{equation}
For frequencies $\omega=\omega_1-\omega_0=\omega_m$, the criterion in equation \eqref{eq:back_action_criterion} is satisfied  if $Q_\text{BA}\gg Q_1Q_m$. Even though increasing $Q_\text{BA}$ can mean lowering the EM-mechanical coupling strength, it can still improve the signal power on resonance by lowering back-action damping. One way to achieve this is by increasing the cavity mass $M$, as this also does not affect the signal power off-resonance\footnote{In order to avoid back-action effects completely by raising the mass, as assumed e.g.\ in \cite{dagnolo2025classicalandquantumheuristics}, unfeasible masses as large as $M\gg 2\cdot 10^{12}\,\text{kg}\,\frac{Q_1}{10^{10}}\frac{Q_m}{10^{6}}\frac{A_S^2}{(1\,\text{m})V}\frac{B_0^2}{(0.1\,\text{T})^2}\frac{\text{kHz}^2}{f_m^2}$ can be required.}. However, the optimal solution is to lower $Q_1$ through overcoupling, which not only increases the signal power on resonance until equation \eqref{eq:back_action_criterion} is fulfilled, but also increases the signal power off-resonance\footnote{The suggestion in references \cite{ballantini_microwave_2005} and \cite{lowenberg_lorentz_2023} is to lower the coupling strength $C_{01}^m$ or the pump amplitude $B_0$ instead. While this can also help to satisfy equation \eqref{eq:back_action_criterion}, it reduces signal power off-resonance and is never more beneficial than overcoupling. However, if sufficient overcoupling cannot be achieved to maximize the signal power or reach the mechanical thermal noise floor, adjusting $C_{01}^m$ or $B_0$ can still help.}. Since the mechanical quality factors can be as large as $Q_m\sim10^6$ \cite{ballantini_microwave_2005}, an unfeasible amount of overcoupling 
\begin{equation}\label{eq:Q_max_r_m_res}
Q_\text{cpl}^\text{max. m. res.}=\frac{Q_\text{BA}}{Q_m}    
\end{equation}
can be required to fulfill equation \eqref{eq:max_m_res_Q_cpl} considering the orders of magnitude in equation \eqref{eq:Q_back_action}. However, there is no gain in GW sensitivity, as soon as the dominant noise source is given by thermal vibrations, which get amplified by the same transfer function as GW-induced vibrations. Hence, another important quantity to consider is the value of $Q_1\approx Q_\text{cpl}$ for which mechanical thermal noise equals thermal readout noise on mechanical resonance
\begin{equation}\label{eq:Q_m_th_r}
    Q_\text{cpl}^\text{m.th.=r}\simeq
        2\frac{Q_\text{BA}}{Q_m}\frac{n_T^m}{n_T^r}\,,
\end{equation}
where we have assumed $Q_\text{cpl}\ll Q_\text{int}$ and neglected a second solution $Q_\text{cpl, 2}^\text{m.th.=r}$ since it is typically too small $Q_\text{cpl, 2}^\text{m.th.=r}\ll1$. Furthermore, we used $n_T^m=k_BT_c/\hbar\omega_m$ to describe the number of thermal photons at the mechanical frequency and used $n_T^m/n_T^r\sim\omega_1/\omega_m\gg1$.
Any detector designed to operate on a mechanical resonance should choose its coupling between the values given in equations \eqref{eq:Q_m_th_r} and \eqref{eq:Q_max_r_m_res}. For example the coupling
\begin{equation}\label{eq:Q_opt_mech_res}
    Q_\text{cpl}^\text{opt.}(\omega_g)=\frac{(\omega_0+\omega_g)\omega_g^2\omega_1}{\left((\omega_0+\omega_g)^2-\omega_1^2\right)\omega_g^2-\gamma_1\gamma_m}
\end{equation}
is usually in between both values and optimizes the signal power at a frequency $\omega_0+\omega_g>\omega_0+\omega_m+\frac{\omega_m}{Q_m}$ while $\omega_1-\omega_0=\omega_m$ still holds. Different choices for this coupling are illustrated in figure \ref{fig:mechanical_resonance_sensitivity} on the top right.

However, no matter how $Q_\text{cpl}$ is adjusted, below $Q_\text{cpl}^\text{m.th.=r}$ the maximal GW strain sensitivity on resonance is determined by the amount of thermal vibrations of the cavity walls which are amplified and damped in the same way as the GW signal
\begin{equation}\label{eq:mech_res_m_th_noise_strain}
    \sqrt{S_{h}^\text{m.\,th.}(\omega_1=\omega_0+\omega_m)}\simeq7\cdot10^{-23}\,\text{Hz}^{-\frac{1}{2}}\,\sqrt{\frac{T_c}{1.8\,\text{K}}\frac{10^3\,\text{kg}}{M}\frac{10^6}{Q_m}\left(\frac{2\pi\cdot 10\,\text{kHz}}{\omega_m}\right)^3}\,\frac{1\,\text{m}}{R_\text{eff}}\frac{1}{\Gamma_m}\,.
\end{equation}
Fortunately, thermal emissions from the cavity walls enter the cavity as a field $b_\text{in}$ in equation \eqref{eq:b_1_solution_with_damp} and get suppressed by the mechanical resonance $\sim Q_m^{-1}$ relative to the signal power (see also figure \ref{fig:mechanical_resonance_sensitivity}). Therefore, thermal emissions can be safely neglected in comparison to thermal vibrations on resonance. 
Overall, the benefit of operating on a mechanical resonance is the possibility to significantly overcouple and hence gain bandwidth, without losing the sensitivity in equation \eqref{eq:mech_res_m_th_noise_strain} on resonance.

\subsection{Off mechanical resonance}
The signal power can also be damped for frequencies $\omega_g=\omega_1-\omega_0\gg\omega_m$ unless 
\begin{equation}\label{eq:damping_off_res_condition}
    Q_1\ll Q_\text{BA} \,.
\end{equation}
Again, $Q_\text{BA}$ can be increased with a large cavity mass and $Q_1$ can be lowered through overcoupling. The coupling that maximizes $T_\text{sig}$ at $\omega_g=\omega_1-\omega_0\gg\omega_m$ is given by
\begin{equation}\label{eq:optimal_beta_off_mech_res}
    Q_\text{cpl}^\text{max. off m. res.}=\left(\frac{1}{Q_\text{int}^2}+\frac{1}{Q_\text{BA}^2}\right)^{-\frac{1}{2}}.
\end{equation}
We can also obtain the coupling which maximizes the signal transfer function for $\omega_1-\omega_0\ll\omega_m$ by exchanging $Q_\text{BA}\to Q_\text{BA}^m\coloneqq\omega_1^2\omega_m^2/\gamma_1\gamma_m$ in equation \eqref{eq:optimal_beta_off_mech_res}.

As before, there is no need to maximize the signal power alone, when another resonant noise source dominates. For $\omega_g=\omega_1-\omega_0\gg\omega_m$, electromagnetic thermal emissions from the cavity walls are usually stronger than noise from thermal vibrations. Again, we can find a $Q_\text{cpl}$ for which readout noise in equation \eqref{eq:readout_noise_PSD} and thermal emissions in equation \eqref{eq:thermal_noise_PSD} are equal using $n_T^c=k_BT_c/\hbar\omega_1$ 
\begin{equation}\label{eq:Q_th_r_off_mech_res}
     Q_\text{cpl}^\text{th=r}=\frac{Q_\text{int}}{2\frac{n_T^c}{n_T^r}-1+\sqrt{\left(2\frac{n_T^c}{n_T^r}\right)^2-4\frac{n_T^c}{n_T^r}-\left(\frac{Q_\text{int}}{Q_\text{BA}}\right)^2}}\xrightarrow{Q_\text{int}/Q_\text{BA}\to0}\begin{cases}
        \frac{Q_\text{int}}{4}\frac{n_T^r}{n_T^c}\,,&\quad n_T^c\gg n_T^r\\Q_\text{int}\,,&\quad n_T^c= n_T^r
    \end{cases}\;,
\end{equation}
where we are only considering the lower of two possible solutions for $Q_\text{cpl}^\text{th=r}$, as it gives a broader sensitivity. Choosing this coupling will maximize the detector's bandwidth while maintaining maximal possible sensitivity on resonance. However, only in case 
\begin{equation}\label{eq:th_r_equality_condition}
2\frac{n_T^c}{n_T^r}\geq1+\sqrt{1+\left(\frac{Q_\text{int}}{Q_\text{BA}}\right)^2}\,,    
\end{equation}
the thermal noise floor can be reached through overcoupling. Otherwise, the SNR with readout noise is maximized using equation \eqref{eq:optimal_beta_off_mech_res}.

\begin{figure}[hb]
    \centering
    \includegraphics[width=0.48\textwidth]{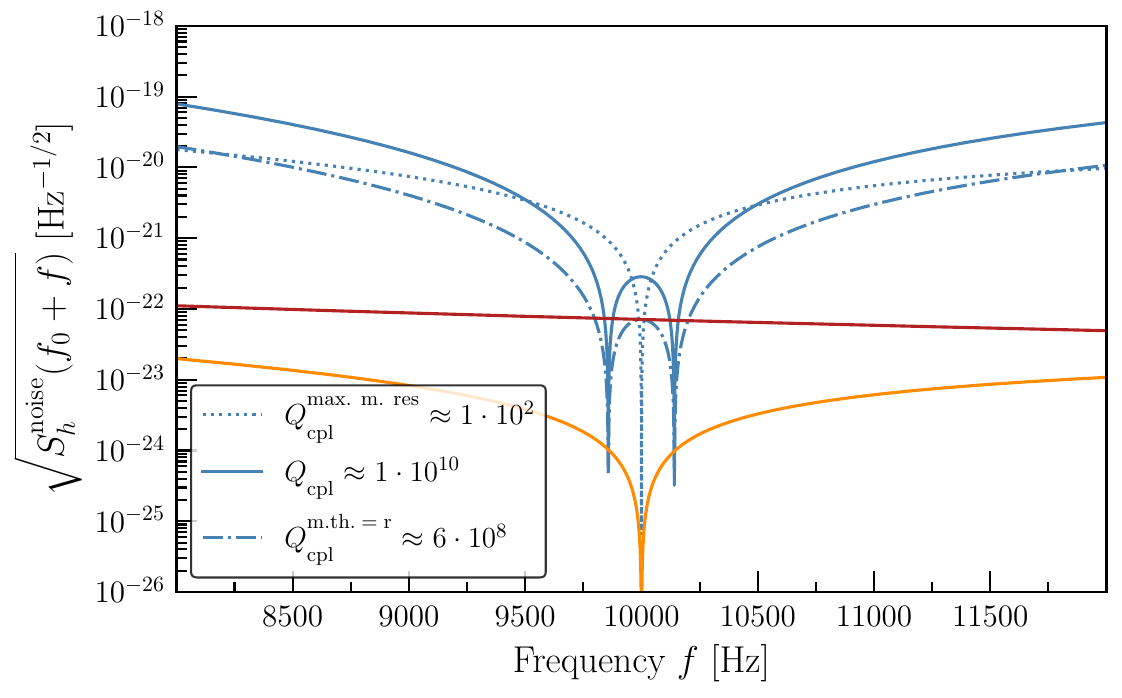}%
    \hfill
    \includegraphics[width=0.48\textwidth]{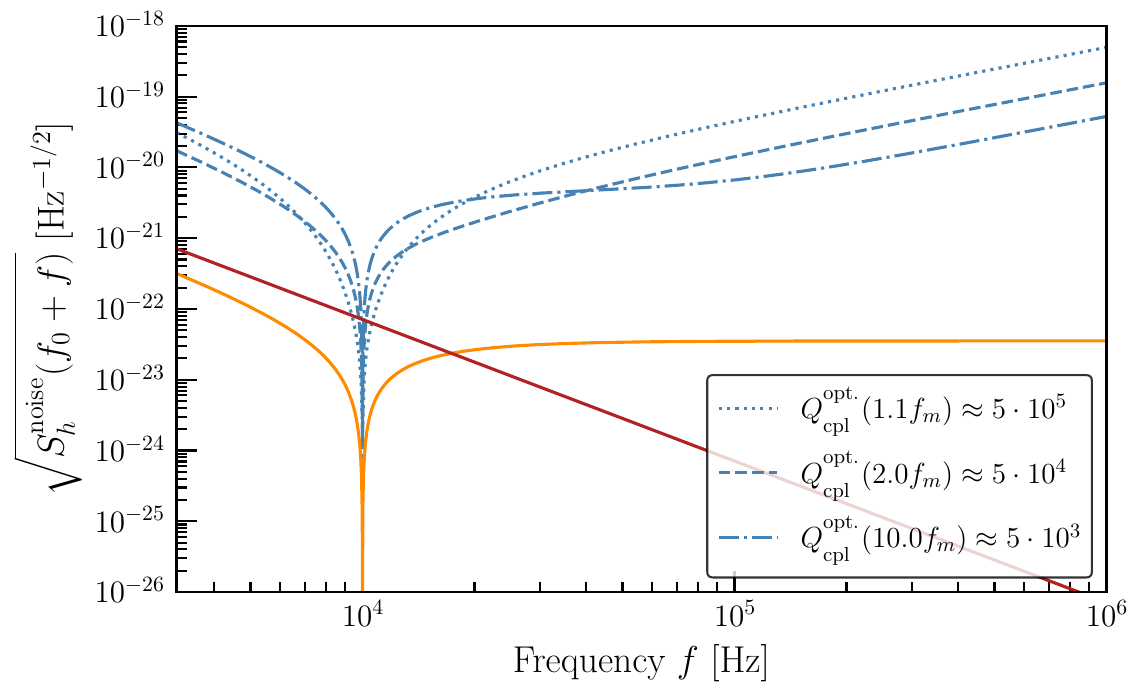}

    \includegraphics[width=0.48\textwidth]{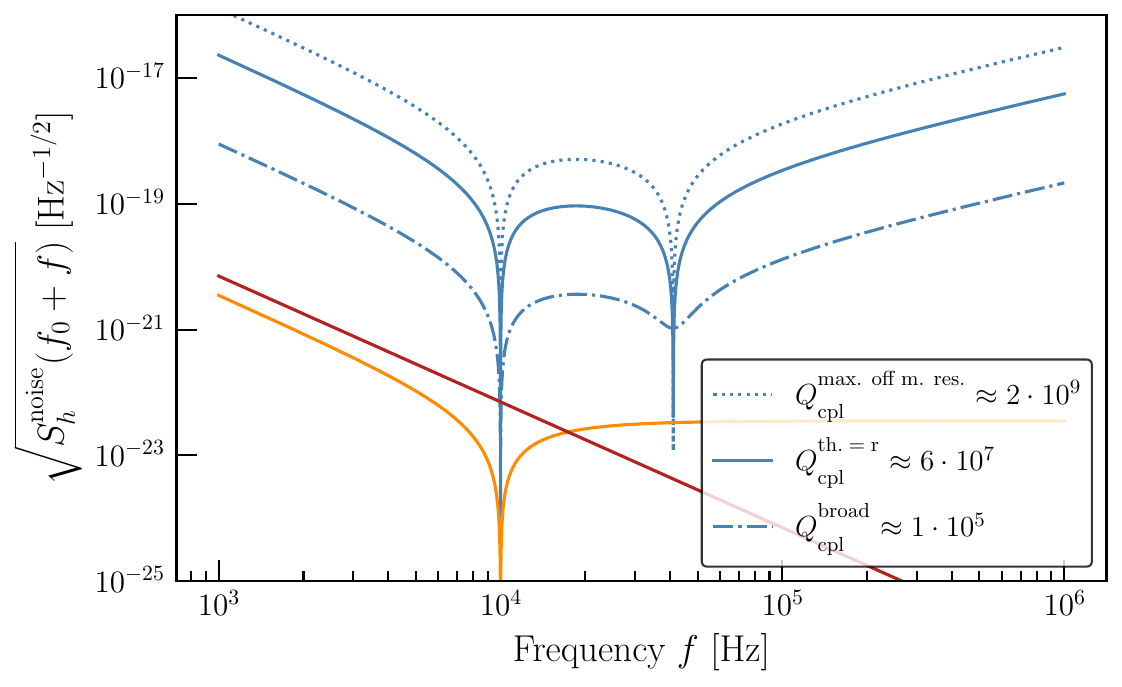}%
    \hfill
    \includegraphics[width=0.48\textwidth]{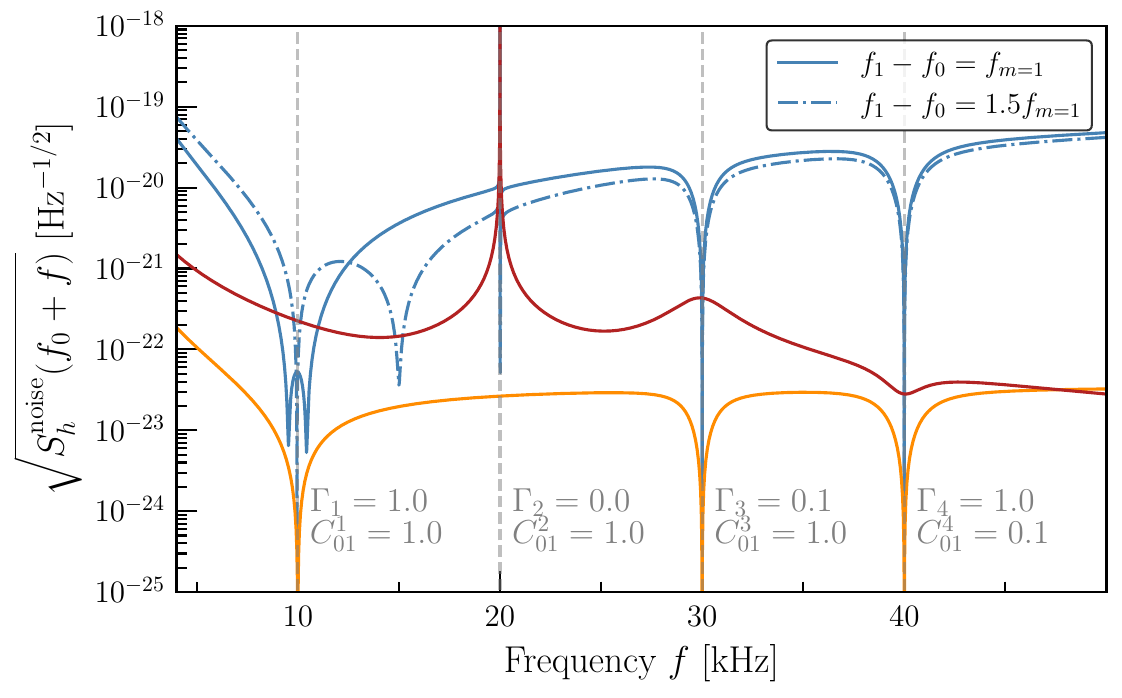}
    \caption{The noise strains of readout noise (blue) for $f_1-f_0=f_m=10\,\text{kHz}$ (top left and right) and $\omega_1-\omega_0>\omega_m+\omega_m/Q_m$ (bottom left) for different choices of $Q_\text{cpl}$ from equations \eqref{eq:max_m_res_Q_cpl}, \eqref{eq:Q_m_th_r}, \eqref{eq:Q_opt_mech_res}, \eqref{eq:optimal_beta_off_mech_res} and \eqref{eq:Q_th_r_off_mech_res}. Thermal radiation from the cavity walls (orange) and thermal vibrations of the cavity (red) are shown for comparison. On the bottom right, the noise strains for multiple mechanical resonances (grey) with different couplings $\Gamma_m$ and $C_{01}^m$ are compared for two values of $\omega_1-\omega_0$  with $Q_\text{cpl}=10^7$.  The remaining parameters are chosen as $f_0=\text{GHz}$, $f_1=f_0+10\,\text{kHz}$, $Q_m=10^6$, $Q_\text{int}=10^{10}$, $V/\text{m}^3=A/\text{m}^2=R_\text{eff}/\text{m}=1$, $\norm{\bm{B}_0}_S=\norm{\bm{B}_1}_S=\underset{\bm{x}\in\partial V}{\max}\bm{B}_0=1$, $M=10^3\,\text{kg}$, $T_c=1.8\,\text{K}$, $n_T^r  = 1$ and $B_0=0.2\,\text{T}$.}
    \label{fig:mechanical_resonance_sensitivity}
\end{figure}

Furthermore, small parameter regions exist in which $\omega_1-\omega_0\gg\omega_m+\omega_m/Q_m$, but thermal mechanical noise still dominates. The dashed curve in the bottom right of figure \ref{fig:mechanical_resonance_sensitivity} shows an example of this. Equality between readout noise and thermal mechanical vibrations is reached for the couplings
\begin{equation}\label{eq:Q_off_m_res_m_th_r_equality}
   Q_\text{cpl}^\text{m.th.=r}= \frac{n_T^m}{n_T^r}\frac{Q_\text{BA}^m}{Q_\text{m}}\pm\sqrt{\left(\frac{n_T^m}{n_T^r}\frac{Q_\text{BA}^m}{Q_\text{m}}\right)^2-\left(Q_\text{BA}^m-Q_\text{BA}\right)^2}\,,
\end{equation}
which is valid if $Q_\text{cpl}^\text{m.th.=r}\ll Q_\text{int}$. Whenever possible, $Q_\text{cpl}$ should be chosen near the lower solution of equation \eqref{eq:Q_off_m_res_m_th_r_equality}, as it leads to a larger bandwidth.

\subsection{Multiple mechanical resonances}
So far, we have only assumed the presence of a single relevant mechanical resonance. In practice, it is expected that multiple modes will have $\Gamma_m\sim1$ and a possibly very different set of modes will have $C_{01}^m\sim1$. The output PSD obtained from solving equations \eqref{eq:EOM_b_with_BA_short} and \eqref{eq:EOM_q_with_BA_short} with multiple mechanical modes is given by
\begin{equation}
    S_\text{out}(\omega)=\frac{1}{2\mu_0}\frac{\omega_1}{Q_\text{cpl}}\frac{V}{|\text{Res}(\omega)|^2}\left[\sum_m\frac{\gamma_1^2/M^2}{|R_m(\omega-\omega_0)|^2}S_{f_\text{ext}^m}(\omega-\omega_0)+\omega_1^4S_{b_\text{in}}(\omega)\right]\,,
\end{equation}
where we have introduced the resonance function $\text{Res}(\omega)=R_1(\omega)-\sum_m\frac{\gamma_1\gamma_m}{R_m(\omega-\omega_0)}$ and assumed that the external force couples incoherently to the different mechanical modes. Even if a mechanical resonance is far away from the EM mode splitting $\omega_1-\omega_0$, it can still affect the signal transfer function, which is shown on the bottom right in figure \ref{fig:mechanical_resonance_sensitivity}. The role each resonance plays depends on the combination of coupling coefficients between GW and the mechanical mode $\Gamma_m$ and between the mechanical mode and the EM modes $C_{01}^m$. We assume in equation \eqref{eq:thermal_mechanical_noise_PSD} that each vibrational mode gets thermally excited independent of its displacement field. Therefore, every mechanical mode with $C_{01}^m\neq0$ will contribute noise to the signal mode. However, only the modes with $\Gamma_m C_{01}^m\neq0$ can contribute a GW signal. Therefore, when designing a cavity that uses mechanical resonances for GW detection, it should ideally \emph{only} allow large $C_{01}^m$ when $\Gamma_m$ is also large which is illustrated in figure \ref{fig:mechanical_resonance_sensitivity} as well.

\section{Figure of merit components}\label{sec:FOMs}
When designing an SRF cavity for gravitational wave detection, choices need to be made about the GW frequencies $\text{kHz}<\omega_g\lesssim\text{GHz}$ it should be sensitive to, whether the mechanical signal should be enhanced by one or multiple resonant vibrations, and whether the detector will be used in a broadband search or by resonantly scanning through frequencies. Detectors optimized for different sets of choices can look entirely different. The goal of this section is to identify the relevant parameters to combine into figures of merit (FOMs) for all common scenarios listed above and provide expressions for the resonant strain sensitivity, optimal choice of antenna coupling, and cavity bandwidth. The main results, which can be combined into FOMs for an optimization, are summarized in tables \ref{tab:FOM_Collection} and \ref{tab:FOM_scaling} and will be explained in the following. As in section \ref{sec:optimizing_the_geometry}, we will distinguish between two GW frequency regimes.

\subsection{Long-wavelength regime}
As we have seen in the previous appendix \ref{sec:mech_res}, the coupling to the signal mode $Q_\text{cpl}$ is a crucial parameter to choose. The benefit is that it can be adjusted over many orders of magnitude within a given cavity by only changing the pickup antenna's position, orientation, or geometry. 

The first case we consider in table \ref{tab:FOM_Collection}, row \ref{R1} is when mechanical vibrations dominate over thermal emissions on resonance. This is true when the detector's EM resonance is in the immediate vicinity of a strongly interacting mechanical resonance $\omega_1-\omega_0\simeq\omega_m$. In this case we need not distinguish between resonant and broadband detection, since we can lower $Q_\text{cpl}$ and hence the bandwidth over a large range $Q_\text{cpl, 2}^\text{m.th=r}< Q_\text{cpl}<Q_\text{cpl}^\text{m.th=r}$ without losing any sensitivity on resonance. We typically have $Q_\text{cpl, 2}^\text{m.th=r}\ll1\ll Q_\text{cpl}^\text{max. m. res.}$, however no values for $Q_\text{cpl}$ lower than $Q_\text{cpl}^\text{max. m. res.}$ are worth considering. This is because the sensitivity bandwidth is mostly determined by the mechanical bandwidth $\omega_m/Q_m$ when the EM bandwidth $\omega_1/Q_1$ is much larger. An example of this can be seen in the dotted line in figure \ref{fig:mechanical_resonance_sensitivity} on the top left. By choosing $Q_\text{cpl}=Q_\text{cpl}^\text{opt.}(\omega_g)$ according to equation \eqref{eq:Q_opt_mech_res}, the off-resonant sensitivity at frequency $\omega_g$ is optimized. 

Row \ref{R2} shows the case where $\omega_1-\omega_0$ is outside any mechanical bandwidth but still dominated by noise from thermal vibrations. Here it is best to overcouple until readout noise matches the thermal noise using equation \eqref{eq:Q_off_m_res_m_th_r_equality}.
The bandwidth in both cases not only depends on $Q_\text{cpl}$ but also on the coupling between the mechanical and electromagnetic oscillators $Q_\text{BA}$. Hence, we simply define the bandwidth as the frequency range for which mechanical thermal noise dominates over readout noise, and hence the $\text{SNR}$ is nearly constant as shown in figure \ref{fig:mechanical_resonance_sensitivity}. Since the $\text{SNR}$ on resonance is solely determined by the ratio between GW tidal force and thermal force in this case, it does notably not depend on any electromagnetic properties of the detector. 

The second important class of frequency ranges is when thermal emissions dominate over thermal vibrations on resonance. While this is typically the case when no vibration is excited on resonance, the interaction with the mechanical oscillator can still affect the sensitivity due to back-action. Now, without the aid of a mechanical resonance, strongly overcoupling to the signal mode will quickly reduce sensitivity on resonance. Therefore, we need to distinguish between resonant scanning setups, where sensitivity on resonance is preferred at the cost of low bandwidth, and broadband searches, where resonant sensitivity is traded in for more bandwidth.

However, even for resonant scanning, no sensitivity is lost as long as the coupling is chosen so that readout noise approximately equals noise from thermal emissions. This is shown in row \ref{R3} and is possible under the condition in equation \eqref{eq:th_r_equality_condition} where the coupling is given by equation \eqref{eq:Q_th_r_off_mech_res}. The additional bandwidth is useful even for a resonant scanning setup, since more frequencies can be covered in one scanning step, which either reduces the total experimental time needed or allows increasing the integration time for one scanning step. 
Since thermal emissions are reduced through large $Q_\text{int}$, the total noise strain reduces with it. While the signal power \eqref{eq:mechanical_signal_PSD} improves when most signal fields are located near the surface $\norm{{\bm{{B}}_1}}_S\sim 1$, the quality factor for internal losses decreases with the same factor, since power is only lost due to fields at the surface. As a result, both contributions cancel in the noise strain and the signal mode geometry only enters through the coefficient $C_{01}^g$.

Row \ref{R4} applies if the thermal occupation of the readout is too large, or back-action effects are too strong, so that the condition in equation \eqref{eq:th_r_equality_condition} is not fulfilled, and the noise strain is purely determined by readout noise. In that case, $Q_\text{cpl}$ needs to be chosen to maximize the signal power according to equation \eqref{eq:optimal_beta_off_mech_res}. When back-action dominates, this is $Q_\text{cpl}=Q_\text{BA}$, and when back action is negligible, critical coupling $Q_\text{cpl}=Q_\text{int}$ is optimal. Since $Q_\text{BA}\propto(\omega_1-\omega_0)^2$, the latter case is the limit of the former for high $\omega_1-\omega_0$ and is shown explicitly in row \ref{R5}. 

Finally, a broadband detector in row \ref{R6} is characterized by choosing $Q_\text{cpl}\ll Q_\text{cpl}^\text{max. off m. res.}$ or $Q_\text{cpl}\ll Q_\text{cpl}^\text{th=r}$ if it exists. This also leads to a readout noise dominated noise strain. However, since the loaded signal quality factor $Q_1\approx Q_\text{cpl}$, all dependence on $Q_\text{int}$ is lost and a detector does not need to be optimized for it.

\subsection{Wavelengths comparable to detector size}
Since mechanical resonances are not expected to affect the signal transfer function when the GW wavelength becomes comparable to the detector size, less modes of operation need to be distinguished. Again, we are considering the case where a detector is optimized to have maximal possible sensitivity on resonance, which means operating at the noise floor given by thermal emissions in the cavity (row \ref{R7} in table \ref{tab:FOM_Collection}). Like for the mechanical signal, the coupling to the signal mode can be increased until the thermal noise equals the readout noise, which is the case for $Q_\text{cpl}^\text{th=r}$ in equation \eqref{eq:Q_th_r_off_mech_res} in the limit $Q_\text{BA}\to\infty$. Important FOMs in this case are increasing the ratio $V/\sqrt{A_S}$ and lowering surface magnetic fields in the signal mode $\norm{\bm{B}_1}_S$.

When the readout is overcoupled to increase the bandwidth, readout noise dominates, $Q_\text{int}$ does not enter the sensitivity, and the minimal GW strain scale with $\sqrt{V}$, which is shown in row \ref{R8} in table \ref{tab:FOM_Collection}.

\subsection{Summary}
Table \ref{tab:FOM_Collection} summarizes all relevant scenarios in which an SRF cavity can operate for GW detection depending on the frequency range and dominant noise source. We also show the recommended coupling to the signal mode $Q_\text{cpl}$ and the inverse equivalent noise strain that can be achieved with it. The bandwidth is listed separately, however any real sensitivity prediction needs to take strain sensitivity \emph{and} bandwidth into account. However, their specific combination depends on the type of GW signal and mode of operation, as discussed in section \ref{sec:optimizing_the_geometry}. Note that the list is not completely exhaustive, since only desirable modes of operation are considered. This includes cases where the optimal coupling can not be reached or further noise sources dominate. 

The scaling of the different sensitivities with all relevant parameters is further shown in table \ref{tab:FOM_scaling}. Many relevant parameters obey strict upper limits which an optimized detector will approximately fulfill. In particular $\Gamma_m\leq1$, $C_{01}^{m/g}\lesssim1$, $\kappa_{01}\lesssim1$, $\underset{\bm{x}\in\partial V}{\max}|\bm{B}_0|\geq1$, $\frac{\norm{\bm{B}_0}_V}{\underset{\bm{x}\in\partial V}{\max}\bm{B}_0}\leq1$. $\norm{\bm{B}_1}_S$ does not technically obey any bound but in practice rarely reaches values much larger than one. $Q_m$ enters when mechanical vibrations dominate, and should always be as large as possible. However theoretically determining it during an optimization procedure is hard, since it needs to take the suspension system of the cavity and the surrounding liquid helium into account. Increasing the signal frequency $\omega_1$ is always beneficial or irrelevant until the surface resistance is dominated by the BCS resistance $R_\text{BCS}(\omega)>R_\text{res}$. Thus, the only geometrical parameters that can in principle increase the sensitivity indefinitely are related to the size of the cavity. The mechanical interaction improves with $R_\text{eff}$ and $A_S$ and can even decrease with total volume $V$ in a broadband search. In contrast, the electromagnetic interaction only ever improves with the volume $V$.

Further improvements in sensitivity can be reached by increasing the mass $M$ through increasing the thickness of the cavity walls. This limits thermal (and other mechanical) vibrations and back-action effects. Ideally, the temperature of the readout system is lowered until quantum fluctuations dominate the readout noise, and the sensitivity does not scale with $T_r$ anymore. The temperature of the cavity itself also should be as low as possible. However reaching the quantum limit at $T_c\ll 1\,\text{K}$ is not feasible for large pump powers. Lowering the surface resistance $R_S$ and increasing the critical magnetic field of the superconductor $B_c$ are the remaining parameters that can be optimized. In particular, increasing $B_c$ linearly improves the sensitivity and might be possible without further drawbacks by coating niobium cavities with thin films of other materials \cite{Posen_2017}. 

To conclude, as soon as an optimized geometry has been found, all that can be done to improve the sensitivity is to increase the size of the detector, or find ways to improve the superconducting properties of the cavity material.

\begin{table}[htbp]
\begin{tabular}{|c|c|c|c|c|c|}
\hline
Range & Condition & $Q_\text{cpl}$ & $1/\sqrt{S_h^\text{noise}(\omega)}$& Bandwidth & Case \\
\hline
\multicolumn{6}{c}{\textbf{Long wavelengths $\omega_gL_\text{cavity}\ll c$}} \\
\hline

\multirow{2}{*}{\pC{2.3cm}{$S_\text{m.th.}(\omega_1)$\\$>S_\text{th.em}(\omega_1)$}} &
\pC{2.25cm}{$\frac{n_T^m}{n_T^r}\gg1$\\$\omega_1-\omega_0=\omega_m$} &
\pC{4.2cm}{$\frac{Q_\text{BA}}{Q_m}<Q_\text{cpl}<Q_\text{cpl}^\text{m.th.=r}$\\$Q_\text{cpl}\simeq Q_\text{cpl}^\text{opt.}(\omega_g)$\;\eqref{eq:Q_opt_mech_res}} &
\pC{4.8cm}{$ \frac{R_\text{eff}\Gamma_m}{4}\sqrt{\frac{M Q_m\omega_m^3}{\pi k_B T_c}}$} &
\pC{2.2cm}{$\omega$: $S_\text{m.th.}(\omega)$\\$>S_\text{r}(\omega)$} &
\refstepcounter{tab_row}(\thetabrow)\label{R1}
\\\cline{2-4}

& 
\pC{2.25cm}{$\omega_1-\omega_0\gg\omega_m$} & 
\pC{4.2cm}{$Q_\text{cpl}^\text{m.th=r}$\,\eqref{eq:Q_off_m_res_m_th_r_equality}} &
\pC{4.8cm}{$\frac{R_\text{eff}\Gamma_m(\omega_1-\omega_0)^2}{4}\sqrt{\frac{M Q_m}{2\pi k_B T_c\omega_m}}$} &
&
\refstepcounter{tab_row}(\thetabrow)\label{R2}
\\
\hline

\multirow{3}{*}{\pC{2.3cm}{$S_\text{th.em}(\omega_1)$\\$>S_\text{m.th.}(\omega_1)$}} &
\pC{2.25cm}{$2\frac{n_T^c}{n_T^r}\geq$\\$\text{max}(\frac{Q_\text{int}}{Q_\text{BA}},2)$} &
\pC{4.2cm}{$Q_\text{cpl}^\text{th=r}\;\eqref{eq:Q_th_r_off_mech_res}$} &
\pC{4.8cm}{$\frac{R_\text{eff}B_c\norm{\bm{B}_0}_SC^g_{01}\omega_1\sqrt{A_S}}{8\sqrt{\pi k_BT_cR_S(\omega_1)}\max_{\partial V}|\bm{B}_0|}$} &
\pC{2.2cm}{$\frac{\omega_1}{Q_\text{cpl}^\text{th=r}}$} &
\refstepcounter{tab_row}(\thetabrow)\label{R3}
\\\cline{2-6}

&
\pC{2.25cm}{$2\frac{n_T^c}{n_T^r}<$\\$\text{max}(\frac{Q_\text{int}}{Q_\text{BA}},2)$} &
\pL{4.2cm}{$Q_\text{cpl}=Q_\text{cpl}^\text{max. off m. res.}$\,\eqref{eq:optimal_beta_off_mech_res}\\$\xrightarrow{Q_\text{BA}\ll Q_\text{int}} Q_\text{BA}$\\$\xrightarrow{Q_\text{int}\ll Q_\text{BA}}Q_\text{int}$} &
\pC{4.8cm}{\quad\\$\frac{\sqrt{M}R_\text{eff}|\omega_1-\omega_0|\Gamma_m}{\sqrt{8\pi\hbar n_T^r}}$\\ $\frac{C_{01}^gR_\text{eff}B_c\norm{\bm{B}_0}_S\sqrt{\omega_1A_S}}{\sqrt{32\pi\hbar n_T^rR_S(\omega_1)}\max_{\partial V}|\bm{B}_0|}$} &
\pC{2.2cm}{{\,}\\$\frac{\omega_1}{Q_\text{BA}}$\\ $\frac{\omega_1}{Q_\text{int}}$} &
\pC{1cm}{{\,}\\\refstepcounter{tab_row}(\thetabrow)\label{R4}\\\refstepcounter{tab_row}(\thetabrow)\label{R5}}
\\\cline{2-6}

&
\pC{2.25cm}{Broadband} &
\pC{4.2cm}{$Q_\text{cpl}\ll Q_\text{cpl}^\text{th=r}$\\ or $Q_\text{cpl}\ll Q_\text{cpl}^\text{max. off m. res.}$} &
\pC{4.8cm}{$\frac{\sqrt{Q_\text{cpl}}A_SB_cR_\text{eff}C_{01}^g\norm{\bm{B}_0}_S\norm{\bm{B}_1}_S}{\sqrt{8\mu_0\pi\hbar n_T^rV}\max_{\partial V}|\bm{B}_0|}$} &
\pC{2.2cm}{$\frac{\omega_1}{Q_\text{cpl}}$} &
\refstepcounter{tab_row}(\thetabrow)\label{R6}
\\\hline

\multicolumn{6}{c}{\textbf{Wavelengths comparable to detector size\footnote{The same figures of merit apply when the electromagnetic signal is caused by an axion effective current, by only exchanging the coupling coefficient $\kappa_{01}$.}}} \\
\hline

\multirow{2}{*}{\pC{2.3cm}{$\omega_gL_\text{cavity}\sim c$\\ $\omega_g\gg\omega_m$}} &
\pC{2.25cm}{Resonant\\ $\frac{n_T^c}{n_T^r}\geq1$} &
\pC{4.2cm}{$Q_\text{cpl}^\text{th=r}$\eqref{eq:Q_th_r_off_mech_res}} &
\pC{4.8cm}{$\frac{V\omega_1|\kappa_{01}|B_c}{8\sqrt{\pi k_BT_c A_SR_S(\omega_1)}\norm{\bm{B}_1}_S\max_{\partial V}|\bm{B}_0|}$} &
\pC{2.2cm}{$\frac{\omega_1}{Q_\text{cpl}^\text{th=r}}$} &
\refstepcounter{tab_row}(\thetabrow)\label{R7}
\\\cline{2-6}

&
\pC{2.25cm}{Broadband} &
\pC{4.2cm}{$Q_\text{cpl}\ll Q_\text{cpl}^\text{th=r}$} &
\pC{4.8cm}{$\frac{\sqrt{Q_\text{cpl}V}B_c|\kappa_{01}|}{\sqrt{8\mu_0\pi\hbar n_T^r}\max_{\partial V}|\bm{B}_0|}$}  &
\pC{2.2cm}{$\frac{\omega_1}{Q_\text{cpl}}$} &
\refstepcounter{tab_row}(\thetabrow)\label{R8}
\\
\hline
\end{tabular}
\caption{Collection of components for figures of merit for different SRF cavity detector setups.}
\label{tab:FOM_Collection}
\end{table}


\begin{table}[htbp]
    \centering
    \begin{tabular}{|c||c|c|c|c|c|c|c|c|c|c|c|c||c|c|c|c|c|}
         \hline
         Case&$R_\text{eff}$&$A_S$&$V$&$\Gamma_m$&$C_{01}^{m/g}$&$\kappa_{01}$&$Q_m$&$\omega_m$&$\omega_1$&$\norm{\bm{B}_0}_S$&$\underset{\bm{x}\in\partial V}{\max}|\bm{B}_0|$&$\norm{\bm{B}_1}_S$&$B_c$&$T_c$&$T_r$&$M$&$R_S(\omega)$  \\\hline\hline
         C\ref{R1}&1& & &1 & & &$\frac12$ &$\frac32$ & & & & 
 & &$-\frac12$& &$\frac12$&
         \\ \hline
         C\ref{R2}&1& & &1 & & &$\frac12$ &$-\frac12$ & & & && &$-\frac12$& &$\frac12$&
         \\ \hline
         C\ref{R3}&1&$\frac12$ & & &1 & & &&$1\to0$ &1 &$-1$ &&1 &$-\frac12$& &&$-\frac12$
         \\ \hline
         C\ref{R4}&1& & &1 & & & & & & & &  & &&$0\to-\frac12$&$\frac12$&
         \\ \hline
         C\ref{R5}&1&$\frac12$ & & &1 & & & &$\frac12\to-\frac12$&1 &$-1$ & &1 & &$0\to-\frac12$&&$-\frac12$
         \\ \hline
         C\ref{R6}&1&1 &$-\frac12$ & &1 & & & & & 1&$-1$ &1 &1&&$0\to-\frac12$&&
         \\ \hline
         C\ref{R7}&&$-\frac12$ &1 & & &1 & & &$1\to0$ & &$-1$ &$-1$&1&$-\frac12$&&&$-\frac12$
         \\ \hline
         C\ref{R8}& & &$\frac12$ & & &1 & & & & &$-1$ &&1&&$0\to-\frac12$&&
         \\ \hline
    \end{tabular}
    \caption{The scaling of all relevant parameters for optimizing SRF cavities for GW detection. Parameters depending on the cavity geometry have been separated on the left side. The cases correspond to the rows in table \ref{tab:FOM_Collection} and the values in the cells denote the power with which the quantity enters $1/\sqrt{S_h^\text{noise}(\omega)}\sim\sqrt{\text{SNR}}$, which should be \emph{maximized}. Empty cells correspond to no dependence on the parameter. The expressions in the $\omega_1$ and $T_r$ columns show the low $\to$ high frequency/temperature limits.}
    \label{tab:FOM_scaling}
\end{table}

\section{Coupling coefficients for orthogonal rectangular cavities}\label{sec:Rectangular_overlaps}
In this section we provide analytical expressions for the GW coupling coefficients to 3D orthogonal rectangular cavities with freely falling walls in the long-wavelength limit. We will use this result to support the claim in section \ref{sec:2D_optimization} that the angular response of such a detector is primarily determined by the orthogonal cavity setup and \emph{not} by the aspect ratio or specific geometry of the individual cells. This justifies optimizing the 2D geometry only for one particular GW incidence angle as we do in section \ref{sec:2D_optimization}. 

The electromagnetic fields in a rectangular cavity resonator can be split into TE$_{mnp}$ and TM$_{mnp}$ modes. For a cavity with lengths $L_x$, $L_y$ and $L_z$, the frequency $\omega_{mnp}$ is
\begin{align}
\omega_{mnp}=ck_\text{tot}=c\sqrt{k_x^2+k_y^2+k_z^2}\,,
\end{align}
where we have defined the wave numbers
\begin{align}
k_x\equiv\frac{\pi m}{L_x},\quad k_y\equiv\frac{\pi n}{L_y},\quad k_z\equiv\frac{\pi p}{L_z}\,,\quad  k_t=\sqrt{k_x^2+k_y^2}\,.
\end{align}
With this, the magnetic fields in the cavity are
\begin{align}
\bm{B}^\text{TM}&=-\frac{i\sqrt{8}}{k_t}\begin{pmatrix}-k_y\,\sin{k_xx}\cos{k_yy}\cos{k_zz} \\k_x\,\cos{k_xx}\sin{k_yy}\cos{k_zz} \\0\end{pmatrix}\,,\\
\bm{B}^\text{TE}&=-\frac{\sqrt{8}}{\omega_{mnp}}\frac{k_z}{k_t}\begin{pmatrix}k_x\,\sin{k_xx}\cos{k_yy}\cos{k_zz} \\ k_y\,\cos{k_xx}\sin{k_yy}\cos{k_zz}\\ -k_t^2/k_z\,\cos{k_xx}\cos{k_yy}\sin{k_zz}\end{pmatrix}\,,
\end{align}
and the electric fields are given by $\bm{E}=\frac{1}{i\omega_{mnp}}\nabla\times\bm{B}$\,.

The dominant coupling to evaluate is given by $\kappa_{01}^{ij}$ in equation \eqref{eq:directCouplingCoefficient} which is equivalent to evaluating equation \eqref{eq:C01_full_mech_sum} in this frequency range. For a single cavity cell with the same pump and signal mode, we find $\kappa_{01}^{11}=\frac{1}{k_\text{tot}^2}\left(k_z^2+k_y^2\right)$, $\kappa_{01}^{22}=\frac{1}{k_\text{tot}^2}\left(k_z^2+k_x^2\right)$, $\kappa_{01}^{33}=\frac{1}{k_\text{tot}^2}\left(k_x^2+k_y^2\right)$, $\kappa_{01}^{i\neq j}=0$.

The average of the coupling in equation \eqref{eq:directCouplingCoefficient} over all GW incidence angles and polarizations, can be expressed as 
\begin{align}
    \langle&\kappa_{01}^2\rangle=\int_0^{2\pi}\frac{d\psi}{2\pi}\int_0^\pi \frac{d\theta\sin{\theta}}{2}\int_0^{2\pi}\frac{d\varphi}{2\pi}\,|\kappa_{01}(\psi,\varphi,\theta)|^2\\
    &=\frac{4}{15}\left[(\kappa_{01}^{11})^2+(\kappa_{01}^{22})^2+(\kappa_{01}^{33})^2-\kappa_{01}^{22}\kappa_{01}^{11}-\kappa_{01}^{33} \kappa_{01}^{11}-\kappa_{01}^{22} \kappa_{01}^{33}+3 \left( (\kappa_{01}^{12})^2+(\kappa_{01}^{13})^2+(\kappa_{01}^{23})^2\right)\right]\,.\nonumber
\end{align}
For the single rectangular cavity, this evaluates to
\begin{equation}
    \langle\kappa_{01}^2\rangle^\text{single}=\frac{4}{15}\left[\kappa_{01}^+(\hat{\bm{z}})^2+\frac{(k_x^2-k_z^2)(k_y^2-k_z^2)}{k_\text{tot}^4}\right]\,,
\end{equation}
where we have plugged in the coupling to a plus  polarized GW along the $\hat{\bm{z}}$ direction $\kappa_{01}^+(\hat{\bm{z}})=\kappa_{01}^{11}-\kappa_{01}^{22}$. Clearly, an optimization of a particular GW direction and polarization does not guarantee an optimized sky average as well.

However, the result changes when considering the symmetric and anti-symmetric oscillation of the same mode in two orthogonal identical cavity cells. In that case we find $\kappa_{01}^{11}=-\kappa_{01}^{22}=\frac{1}{2k_\text{tot}^2}\left(k_y^2-k_x^2\right)$, $\kappa_{01}^{33}=\kappa_{01}^{i\neq j}=0$ and the average
\begin{equation}
\langle\kappa_{01}^2\rangle^\text{double}=\frac{1}{5}\kappa_{01}^+(\hat{\bm{z}})^2\,,
\end{equation}
where $\kappa_{01}^+(\hat{\bm{z}})$ is the same as for the single cavity cell.
We can see that the sky average is determined entirely by the response to the optimal incidence angle and polarization. Thus, optimizing for this coupling alone simultaneously maximizes the response to all possible incoming GWs. 

While this result can not be generalized exactly to more advanced cavity geometries, we find that it remains true approximately. This provides a further argument in favour of setting up two orthogonal cavities for GW detection, as it improves the angular response pattern and can easily be optimized by considering only one GW direction.

\bibliography{references}

\end{document}